\documentclass[fleqn,reqno,11pt,a4paper,final]{amsart}
\usepackage[a4paper,left=35mm,right=35mm,top=30mm,bottom=30mm,marginpar=25mm]{geometry} 
\usepackage{amsmath}
\usepackage{amssymb}
\usepackage{amsthm}
\usepackage{amscd}
\usepackage[ansinew]{inputenc}
\usepackage{cite}
\usepackage{bbm}
\usepackage{color}
\usepackage[english=american]{csquotes}
\usepackage[final]{graphicx}
\usepackage{hyperref}
\usepackage{calc}
\usepackage{mathptmx}
\usepackage{bm}
\usepackage{enumerate}

\numberwithin{equation}{section}

\newtheoremstyle{thmlemcorr}{10pt}{10pt}{\itshape}{}{\bfseries}{.}{10pt}{{\thmname{#1}\thmnumber{ #2}\thmnote{ (#3)}}}
\newtheoremstyle{thmlemcorr*}{10pt}{10pt}{\itshape}{}{\bfseries}{.}\newline{{\thmname{#1}\thmnumber{ #2}\thmnote{ (#3)}}}
\newtheoremstyle{remexample}{10pt}{10pt}{}{}{\bfseries}{.}{10pt}{{\thmname{#1}\thmnumber{ #2}\thmnote{ (#3)}}}

\theoremstyle{thmlemcorr}
\newtheorem{theorem}{Theorem}
\numberwithin{theorem}{section}

\theoremstyle{thmlemcorr*}
\newtheorem{theorem*}{Theorem}
\newtheorem{lemma*}[theorem]{Lemma}
\newtheorem{corollary*}[theorem]{Corollary}
\newtheorem{proposition*}[theorem]{Proposition}
\newtheorem{problem*}[theorem]{Problem}
\newtheorem{conjecture*}[theorem]{Conjecture}
\newtheorem{definition*}[theorem]{Definition}

\theoremstyle{remexample}
\newtheorem{remark}[theorem]{Remark}
\newtheorem{example}[theorem]{Example}

\newcommand{\Lrm}{\mathrm{L}}

\newcommand{\Nrm}{\mathrm{N}}

\newcommand{\Srm}{\mathrm{S}}
\newcommand{\Trm}{\mathrm{T}}

\newcommand{\Dcal}{\mathcal{D}}
\newcommand{\Ecal}{\mathcal{E}}

\newcommand{\Hcal}{\mathcal{H}}
\newcommand{\Ical}{\mathcal{I}}

\newcommand{\Lcal}{\mathcal{L}}

\newcommand{\Pcal}{\mathcal{P}}

\newcommand{\Scal}{\mathcal{S}}

\newcommand{\Wcal}{\mathcal{W}}

\newcommand{\Hfrak}{\mathfrak{H}}

\newcommand{\Pfrak}{\mathfrak{P}}

\newcommand{\Sfrak}{\mathfrak{S}}

\newcommand{\Zfrak}{\mathfrak{Z}}

\newcommand{\hfrak}{\mathfrak{h}}

\newcommand{\pfrak}{\mathfrak{p}}

\newcommand{\zfrak}{\mathfrak{z}}

\newcommand{\Rbb}{\mathbb{R}}

\DeclareMathOperator{\id}{id}

\DeclareMathOperator{\diverg}{div}
\DeclareMathOperator{\Diverg}{Div}
\DeclareMathOperator{\curl}{curl}
\DeclareMathOperator{\dist}{dist}

\DeclareMathOperator{\tr}{tr}

\DeclareMathOperator{\diag}{diag}
\DeclareMathOperator{\Ad}{Ad}

\DeclareMathOperator{\dev}{dev}

\DeclareMathOperator{\infc}{\square}

\newcommand{\set}[2]{\left\{\, #1 \ \ \textup{\textbf{:}}\ \ #2 \,\right\}}
\newcommand{\setn}[2]{\{\, #1 \ \ \textup{\textbf{:}}\ \ #2 \,\}}
\newcommand{\setb}[2]{\bigl\{\, #1 \ \ \textup{\textbf{:}}\ \ #2 \,\bigr\}}
\newcommand{\setB}[2]{\Bigl\{\, #1 \ \ \textup{\textbf{:}}\ \ #2 \,\Bigr\}}

\newcommand{\setBBB}[2]{\Biggl\{\, #1 \ \ \textup{\textbf{:}}\ \ #2 \,\Biggr\}}
\newcommand{\norm}[1]{\|#1\|}

\newcommand{\abs}[1]{|#1|}

\newcommand{\absb}[1]{\bigl|#1\bigr|}

\newcommand{\absBB}[1]{\biggl|#1\biggr|}

\newcommand{\altnorm}[1]{{\left\vert\kern-0.25ex\left\vert\kern-0.25ex\left\vert #1 \right\vert\kern-0.25ex\right\vert\kern-0.25ex\right\vert}}

\newcommand{\di}{\mathrm{d}}
\newcommand{\dd}{\;\mathrm{d}}
\newcommand{\DD}{\mathrm{D}}
\newcommand{\N}{\mathbb{N}}
\newcommand{\R}{\mathbb{R}}

\newcommand{\start}{\mathrm{start}}

\newcommand{\toup}{\uparrow}
\newcommand{\todown}{\downarrow}

\newcommand{\sbullet}{\begin{picture}(1,1)(-0.5,-2.5)\circle*{2}\end{picture}}
\newcommand{\frarg}{\,\sbullet\,}

\newcommand{\eps}{\epsilon}

\DeclareMathOperator{\Lie}{Lie}

\DeclareMathOperator{\Exp}{Exp}

\DeclareMathOperator{\Diss}{Diss}

\newcommand{\term}[1]{\textbf{#1}}

\newcommand{\GL}{\mathrm{GL}}
\newcommand{\SO}{\mathrm{SO}}
\newcommand{\SL}{\mathrm{SL}}

\newcounter{assumption}
\makeatletter
\newcommand{\nextas}[1]{%
   \refstepcounter{assumption}%
   \protected@write \@auxout{}{\string\newlabel{#1}{{(A\theassumption)}{\thepage}{(A\theassumption)}{#1}{}}}%
   \hypertarget{#1}{(A\theassumption)}%
}
\newcommand{\nextasnamed}[2]{%
   \refstepcounter{assumption}%
   \protected@write \@auxout{}{\string\newlabel{#1}{{(#2)}{\thepage}{(#2)}{#1}{}}}%
   \hypertarget{#1}{(#2)}%
}
\makeatother

\def\XXint#1#2#3{{\setbox0=\hbox{$#1{#2#3}{\int}$} 
\vcenter{\hbox{$#2#3$}}\kern-.5\wd0}}

\newcommand{\restrict}{\begin{picture}(10,8)\put(2,0){\line(0,1){7}}\put(1.8,0){\line(1,0){7}}\end{picture}}

\renewcommand{\eps}{\varepsilon}
\renewcommand{\epsilon}{\varepsilon}
\renewcommand{\phi}{\varphi}

\begin{document}


\title{A two-speed model for finite-strain elasto-plasticity}

\author{Filip Rindler}
\address{\textit{Filip Rindler:} Mathematics Institute, University of Warwick, Coventry CV4 7AL, United Kingdom.}
\email{F.Rindler@warwick.ac.uk}


\hypersetup{
  pdfauthor = {Filip Rindler (University of Warwick)},
  pdftitle = {A two-speed model for finite-strain elasto-plasticity},
  pdfsubject = {},
  pdfkeywords = {}
}


\maketitle


\begin{abstract}
This work presents a new modeling approach to macroscopic, polycrystalline elasto-plasticity starting from first principles and a few well-defined structural assumptions, incorporating the mildly rate-dependent (viscous) nature of plastic flow and the microscopic origins of plastic deformations. For the global dynamics, we start from a two-stage time-stepping scheme, expressing the fact that in most real materials plastic flow is much slower than elastic deformations, and then perform a detailed analysis of the slow-loading limit passage. In this limit, a rate-independent evolution can be expected, but this brings with it the possibility of jumps (relative to the \enquote{slow} time). Traditionally, the dynamics on the jump transients often remain unspecified, which leads to ambiguity and deficiencies in the energy balance. In order to remedy this, the present approach precisely describes the energetics on the jump transients as the limit of the rate-dependent evolutions at \enquote{singular points}. It turns out that rate-dependent behavior may (but does not have to) prevail on the jump transients. Based on this, we introduce the new solution concept of \enquote{two-speed solutions} to the elasto-plastic evolutionary system, which incorporates a \enquote{slow} and a \enquote{fast} time scale, the latter of which parametrizes the jump transients.
\vspace{4pt}


\vspace{4pt}


\vspace{4pt}

\noindent\textsc{Date:} \today{} (version 1.0).
\end{abstract}


\section{Introduction}


The theoretical modelling of large-strain elasto-plasticity for polycrystalline materials poses many challenges and, despite its great importance, no fully satisfactory mathematical theory has emerged so far. Works in this direction include for instance~\cite{GreenNaghdi71,Rice71,Mandel73,NematNasser79,Dafalias87,LubardaLee81,Naghdi90,Zbib93,OrtizRepetto99,Mielke03a,FrancfortMielke06} as well as the monographs~\cite{LemaitreChaboche90,Lubliner08,GurtinFriedAnand10,HanReddy13}. Some major issues in the quest for such a theory are the following: First, elasto-plasticity naturally goes beyond what can be modeled in a traditional continuum mechanics framework. Rather, a body undergoing plastic flow does \emph{not} remain a continuum as the material is internally \enquote{ripped and torn}, even if this is not macroscopically observable. Even if one is not interested in precisely describing the microscopic origins of plastic flow (such as dislocation movement in metals, see for instance~\cite{Kroner01,ContiTheil05,HochrainerZaiserGumbsch07,AbbaschianReedHill09,SandfeldEtAl11,HochrainerEtAl14}), some aspects of the microscopic situation need to be taken into account in order to derive a consistent theory. More advanced mathematical structures such as Lie groups and Lie algebras, can be very effective in describing the relationships between macroscopic and microscopic phenomena, but are rarely used in the engineering literature (but see~\cite{Mielke03a}).

Second, when deriving the full dynamics of the material, it is often unclear how the solid-like elastic deformations and the fluid-like plastic deformations interact. In particular, their relative speed plays an important role, but this does not seem to be used at present for plasticity modelling.

Third, elasto-plastic flow is rate-dependent (viscous) in reality, but only slightly so, and thus rate-independent (quasi-static) approximations are used more often than not. However, this simplification creates the serious problem that the regularity of solution processes can only be low in general since \emph{jumps} (relative to the \enquote{slow} time scale) can occur in rate-independent processes -- at least there is no obvious mechanism preventing this \emph{a-priori}. This creates the need to specify the behavior of the evolution on the jump transients, which is important for the global energetics, but rarely considered.


The present work develops, from first principles, a model of macroscopic elasto-plasticity that aims to addresses the above issues, and then to analyze it. Our approach is based on the following key principles:
\begin{enumerate}
\item Macroscopic deformations are modeled as driven by microscopic slips (e.g.\ slips induced by dislocation movement), an idea taken from single-crystal plasticity. This idea can be abstractly expressed using the theory of Lie groups and their Lie algebras: A matrix from a Lie group represents the macroscopic plastic distortion and hence (part of) the internal state of the material. The plastic flow, however, is specified on the level of the associated \enquote{microscopic} Lie algebra (i.e.\ the tangent space at the identity matrix) as the sum of microscopic \emph{drifts} (slip rates) acting as \emph{infinitesimal generators} of the flow. The main advantage of the Lie-theoretic point of view expressed in this work is that on the microscopic Lie algebra level we are dealing with a \emph{linear} structure.
\item Elastic movements are assumed to be infinitely fast relative to the plastic ones, which is quite realistic~\cite{ArmstrongArnoldZerilli09,BenDavidEtAl14}, so we base the model on the postulate that the system minimizes over all admissible \emph{elastic}, but not plastic, deformations.
\item  The plastic dynamics are first modeled as \emph{rate-dependent}, i.e.\ \emph{viscous}. If the system is spontaneously pushed out of a stable (rest) state, it \emph{relaxes} to stability by following an evolutionary flow rule. Unlike other models (see below for comparisons), here we do not rely on minimization over irreversible movements, which is thermodynamically questionable. In particular, we work with the realistic \emph{local} stability (yield) condition in the slow-loading limit.
\item The combined elasto-plastic dynamics are modeled based on a two-stage time-stepping scheme, which alternates between two \enquote{fundamental motions}: purely elastic minimization and elasto-plastic relaxation, the latter exchanging elastic for plastic distortion without modifying the total deformation. This two-stage approach allows for a very clean modeling without any ambiguity as to which test functions should be considered in the principle of virtual power.
\item Via a slow-loading limit passage we arrive at the limit \enquote{two-speed} formulation, which incorporates two time scales: With respect to the \enquote{slow} time, the formulation is rate-\emph{independent}. On jump transients, however, we retain the possibility of rate-dependent evolution (or a mixture of rate-dependent and rate-independent evolution) with respect to the \enquote{fast} time.
\item We formulate the whole model in the reference frame, but point out how the common formulation with structural (intermediate) tensors is essentially equivalent via Lie group adjoints. We do not use the idea of an \enquote{intermediate (structural) space} since one cannot consistently define \enquote{intermediate points}.
\end{enumerate}

If $y \colon \Omega \subset \R^3 \to \R^3$ denotes the total deformation of our specimen, then the commonly used Kr\"{o}ner--Lee decomposition
\[
  \nabla y = F = E P
\]
splits the deformation gradient into elastic and plastic \emph{distortions} $E, P$; since $E, P$ are not in general curl-free, they might not be deformation gradients themselves. We refer to~
\cite{Kroner60,LeeLiu67FSEP,Lee69EPDF,GreenNaghdi71,CaseyNaghdi80,GurtinFriedAnand10,ReinaConti14} for justifications and various other aspects of this decomposition. Note in particular that if $\curl P \neq 0$, then $E$ cannot be the identity map; this expresses the physical constraint that the elastic deformation has to close the gaps opened by the plastic flow so as to restore a macroscopic continuum.

The multiplicative Kr\"{o}ner--Lee decomposition is at the root of many mathematical challenges in large-strain theories of elasto-plasticity. Not only is it incompatible with our traditional linear function spaces, the splitting of $\nabla y$ into $E$ and $P$ furthermore is clearly not unique. The ensuing ambiguity (often called the \enquote{uniqueness problem} in the literature) is a big obstacle when trying to develop a useful mathematical theory. For instance, if for the moment we consider the macroscopic elasto-plastic flow to be divided into a number of time intervals $[t_k,t_{k+1})$, $k = 0, \ldots, N-1$, then in every such interval we have potentially both an elastic and a plastic distortion, $E_k, P_k$ say. Macroscopically, the total plastic distortion should be $P = P_N P_{N-1} \cdots P_0$. However, if we let $N \to \infty$ (the interval size going to zero), even this \enquote{natural} setup seems to lead to the need for infinite products, which is not feasible due to the non-commutativity of matrix multiplication. For example, an approach using matrix logarithms to transform products into sums runs into trouble since the involved matrices are not necessarily symmetric and positive definite, whereby the matrix logarithm might not be uniquely definable. We here posit that the plastic distortion $P$ can only be determined from the internal state of the material and its flow in time needs to be specified through a differential equation, see~\cite{GurtinFriedAnand10} some discussion on this point. This approach removes all ambiguity in the Kr\"{o}ner--Lee decomposition.

Another feature of the present model, reminiscent of recent work by Reina \& Conti~\cite{ReinaConti14}, is that we study the structure of microscopic slips via a reasoning with functions of bounded variation (see~\cite{EvansGariepy92,AmbrosioFuscoPallara00}) whose derivative contains an absolute continuous (elastic) part and a singular (plastic) part. On the microscopic level these two parts must decompose \emph{additively} since they take place in different parts of the material (bulk and surface parts, respectively). In the macroscopic \enquote{homogenized} theory, however, this separation is lost and we keep track of the plastic distortion rate (speed) via an equation in the \enquote{microscopic} Lie group.

Previous mathematical theories for large-strain elasto-plasticity seem to fall into one of two categories: The first, \emph{energetic}, approach was introduced in~\cite{OrtizRepetto99,MielkeTheil99,MielkeTheilLevitas02,MielkeTheil04} and applied to nonlinear elasto-plasticity in~\cite{Mielke03a,MainikMielke05,FrancfortMielke06,MielkeMuller06,MainikMielke09}. It starts from a time discretization and assumes that the system at every time step minimizes the potential energy plus any dissipational cost that may be incurred by jumping to the target state. In the limit, a global form of the stability (yield) condition and an energy balance can be derived (these two conditions alone, however, can have more solutions than the original time-discrete scheme, see~\cite{MielkeRindler09}). The global, dissipative minimization in the time-stepping scheme assumes infinite foresight of the system since one may jump to an (energetically) far-away state, even if there is a potential energy barrier in the way, so the system can in fact jump \enquote{too early} and \enquote{too far}. Moreover, from a physical perspective, minimization over \emph{irreversible} (dissipative) movements seems to be a violation of the \emph{Second Law of Thermodynamics}. Nevertheless, if these assumptions are a good approximation to the physical situation at hand, then a very mature theory is available; the current state-of-the-art is presented in the recent monograph~\cite{MielkeRoubicek15book}, also see~\cite{Stefanelli08,Stefanelli09} for a global variational principles in this context.

A more recent approach to address the \emph{under-specification} of the system's behavior on jump transients is to add a \emph{vanishing viscosity} term~\cite{OrtizRepetto99,Mielke02,Mielke03a,MielkeRossiSavare09,DalMasoDeSimoneSolombrino10,DalMasoDeSimoneSolombrino11,MielkeRossiSavare12}. Besides the question of what shape of viscosity one should use (which, however, sometimes turns out to be unimportant), the mathematical analysis here is still unfinished and only some special cases of elasto-plastic evolutions can be fully analyzed, usually without infinite-dimensional elastic variables.

Our approximation scheme is closer in spirit to the vanishing viscosity approach and in some cases could be equivalent. However, we try to argue from first principles and only using a time rescaling together with the postulates outlined above.

This paper is organized as follows: After a detailed explanation of the modeling in Sections~\ref{sc:kinematics}--\ref{sc:evolution}, we then in Section~\ref{sc:limit} embark on a detailed, yet mathematically non-rigorous, investigation into the slow-loading limit passage. These calculations shed some light on the total energetic/dissipative behavior of the system and form the basis of a full mathematically rigorous analysis. Such an analysis is the subject of future work, but at present many formidable technical challenges remain.

\section{Kinematics} \label{sc:kinematics}

We start at the microscopic level and with the non-continuum origins of plastic deformations. Then we consider the macroscopic kinematics, which are linked to the microscopic picture through some basic Lie group theory.

\subsection{Microscopic plastic deformations} \label{sec:micro}

\begin{figure}[tb]
\begin{center}
\includegraphics[scale=0.9]{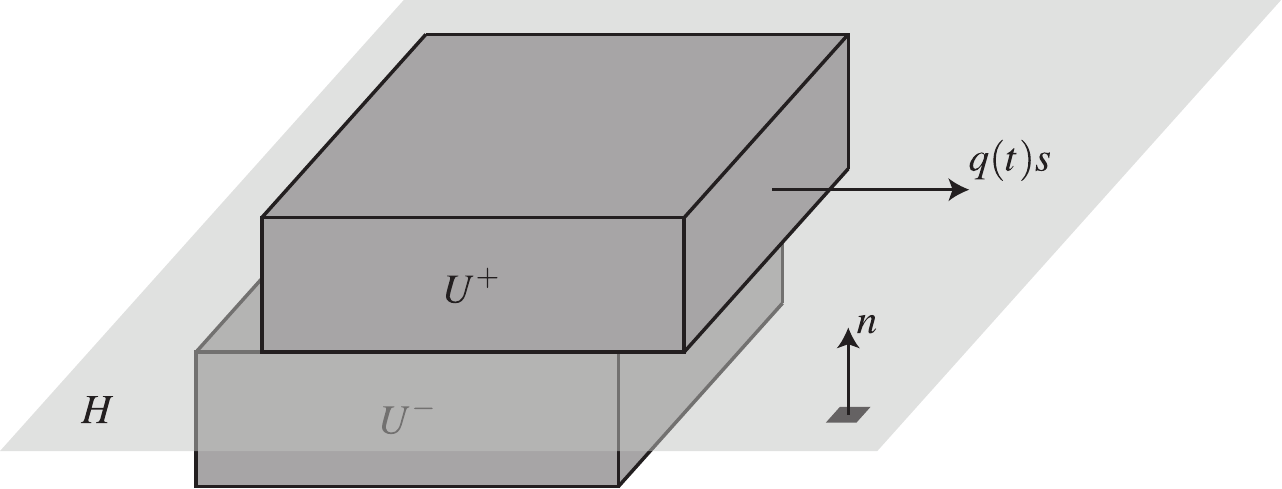}
\caption{A microscopic slip.} 
\label{fig:microslip}
\end{center}
\end{figure}

Consider an open and bounded \enquote{microscopic} reference domain $U \subset \R^d$, $d \in \N$, undergoing plastic deformation. Assume without loss of generality that $0 \in U$ and that there is a (restricted) hyperplane, $H = \set{ x }{ x \cdot n = 0 } \cap U$, defined through its normal vector $n \in \R^d$, $\abs{n} = 1$, that splits $U$ into the two parts
\[
  U^+ := \setn{ x \in U }{ x \cdot n > 0 }, \qquad
  U^- := \setn{ x \in U }{ x \cdot n < 0 }.
\]
We assume that the microscopic plastic deformation manifests itself through a translation of $U^+$ in a referential (relative to the undeformed material) direction $s \in \R^d$ with $\abs{s} = 1$ and $s \perp n$ ($s$ perpendicular to $n$) and with speed $q(t)$ at time $t \geq 0$. The situation is illustrated in Figure~\ref{fig:microslip}.

Hence, the corresponding \term{simple slip} $\gamma_t \colon U \to \R^d$ expressing the deformation after time $t \geq 0$ is
\[
  \gamma_t(x) = \begin{cases}
    x    & \text{if $x \in U^-$,}\\
    x + \biggl( \displaystyle\int_0^t q(\tau) \dd \tau \biggr) s
      & \text{if $x \in U^+$.}
  \end{cases}
\]
If such a simple slip was to occur in a macroscopic state, it would be called a \emph{shear band}, i.e.\ a shear motion over an infinitely thin plane, but we here assume it takes place on a microscopic scale and might not be visible macroscopically.

The space derivative of $\gamma_t$ in the BV-sense is the matrix-valued \emph{measure}
\[
  \DD \gamma_t = I \, \Lcal^d + \biggl(\int_0^t q(s) \dd s \biggr) s \otimes n \, \Hcal^{d-1} \restrict H,
\]
where $\Lcal^d$ is the $d$-dimensional Lebesgue measure (the ordinary $d$-dimensional volume) and $\Hcal^{d-1} \restrict H$ denotes the $(d-1)$-dimensional Hausdorff measure (the $(d-1)$-dimensional area) on $H$ and $I$ is the $d \times d$ identity matrix. Consequently, $\gamma_t$ is a \term{(special) function of bounded variation}. We refer to~\cite{EvansGariepy92,AmbrosioFuscoPallara00} for information on this important class of functions and to~\cite{ReinaConti14} for a justification of the Kr\"{o}ner--Lee decomposition based on a similar argument.

A \enquote{differentiated} view on this simple slip is the following: For the \term{plastic distortion} $P(t) = \DD \gamma_t$ (here there is no elastic distortion), we may derive the following differential equation:
\[
  \left\{ \begin{aligned}
    \dot{P}(t) &= P(t) \bigl[ q(t) s \otimes n \, \Hcal^{d-1} \restrict H \bigr],
      \qquad t > 0, \\
    P(0) &= I,
  \end{aligned}\right.
\]
which we understand as
\[
  \left\{\begin{aligned}
    \dot{P}(t,x) &= 0        &&\text{for $\Lcal^d$-almost every $x \in \Omega$ and $t > 0$,}\\
    \dot{P}(t,x) &= P(t,x) [q(t) s \otimes n]
      &&\text{$(\Hcal^{d-1} \restrict H)$-almost every $x \in \Omega$ and $t > 0$,} \\
    P(0,x) &= I              &&\text{$(\Lcal^d + \Hcal^{d-1} \restrict H)$-almost every $x \in \Omega$.}  \end{aligned} \right.
\]
In particular, at every time $t > 0$, a material vector $m$ from the tangent space to $U$ at a point in $H$ is changed with the rate
\[
  \dot{P}(t) m = q(t) (n \cdot m) P(t) s.
\]
Notice that the referential vector $s$ is transformed into a \enquote{structural} vector $P(t)s$. This corresponds to the assumption that the slip direction $s$ is given with respect to the material frame, which is also natural since the hyperplane $H$ is specified in the material frame (there are no \enquote{structural points} as detailed below) and we need $s \perp n$. However, it is also possible to construct a \enquote{structural} formulation, but this leads to an equivalent theory, see Section~\ref{ssc:ref_struct}.

The above linear matrix differential equation can be solved explicitly using the matrix exponential function (notice that in this special case all \enquote{generator} matrices $q(t) s \otimes n$ commute and $(s \otimes n)^2=0$ since $s \perp n$):
\begin{align*}
  P(t) &= \Exp (0) \, \Lcal^d + \Exp \biggl( \int_0^t q(\tau) s \otimes n \dd \tau \biggr) \, \Hcal^{d-1} \restrict H \\
  &= I \, \Lcal^d + \biggr(\int_0^t q(\tau) \dd \tau \biggl) s \otimes n \, \Hcal^{d-1} \restrict H \\
  &= D\gamma_t,
\end{align*}
as expected.

One can generalize the preceding discussion as follows: We postulate that also in the general situation, the plastic distortion $P(t)$ at time $t$ is given via the \term{plastic distortion equation}
\begin{equation} \label{eq:Fp_flow}
  \dot{P} = P D,
\end{equation}
where $D$ is a $(d \times d)$-matrix field (or even a measure) called the \term{(referential) plastic drift}. In the case of the simple slip above,
\[
  D(t) = q(t) \, s \otimes n \, \Hcal^{d-1} \restrict H.
\]
In metals undergoing plastic distortion, $D$ arises from (activated) crystal defects that move around in the material, causing slip. Since $D$ by definition is a matrix in the \emph{referential} frame, the equation~\eqref{eq:Fp_flow} simply expresses that the referential drift, given by $D$, transformed to the plastically distorted configuration is equal to the change in the latter.

\begin{example}
Often, one considers only plastic drifts $D$ that are a superposition of simple slips (for example all activated slip systems in crystal plasticity). All these slips have trace-free generators ($s \otimes n$ with $s \perp n$) and so in this case we require $D$ to be \emph{deviatoric}:
\[
  \tr\, D = 0.
\]
From the general formula $\partial_t(\det\,A) = \det\,A \cdot \tr\,(A^{-1} \dot{A})$ and~\eqref{eq:Fp_flow}, we get
\[
  \partial_t (\det\,P) = \det\,P \cdot \tr\,D = 0,
\]
hence in this case the \term{plastic incompressibility}
\[
  \det\,P(t) = \det\,P(0) = 1
\]
holds.
\end{example}

It should be remarked that if $D$ is a sum of constant-in-time drifts (e.g.\ several different simple slips), $D = D_1 + \cdots D_n$, then the evolution of $P$ is given as
\[
  P(t) = \Exp \bigl( (D_1 + \cdots D_n)t \bigr) P(0).
\]
This, is however \emph{not} equal to $\Exp(D_1 t) \cdots \Exp(D_n t) P(0)$, because in general the matrices $D_1, \ldots, D_n$ do not commute. To compute this expression, one would need to expand using the Baker--Campbell--Hausdorff formula, which involves the commutator brackets $[D_k,D_l]$ ($k,l = 1,\ldots,n$), see for example Chapter~3 in~\cite{Hall03}. This simple observation already shows that due to the \emph{non-commutativity} of drifts, we need to employ Lie group theory to make sense of the evolution of $P$.

If, as is common in much of the engineering literature (see for instance~\cite{GurtinFriedAnand10}), we instead worked with the \term{structural drift} $L = \dot{P} P^{-1}$, from~\eqref{eq:Fp_flow} we have that
\[
  L = P D P^{-1},
\]
so our $D$ is just the corresponding referential tensor for the structural tensor $L$.


\subsection{Macroscopic kinematics}

We now move from the microscopic to the macroscopic picture. Starting with an open, bounded reference configuration $\Omega \subset \R^d$ of material, we denote the \term{referential (Lagrange, material) points} in it as $x = (x^1,\ldots,x^d) \in \Omega$. Under an elasto-plastic deformation $y = (y^1,\ldots,y^d) \colon \Omega \to \R^d$, every referential point $x$ is mapped to a \term{spatial (Euler) point} $y = y(t,x) \in y(t,\Omega)$, where the time $t$ is from an interval $[0,T)$. A fundamental modeling assumption in this work (as in all of the continuum theory of plasticity) is the following:
\begin{quotation}
  \textit{Macroscopically, the elastically and plastically deformed body is again a continuum without interpenetration of matter or holes; thus $y(t) = y(t,\frarg)$ is a homeomorphism, that is, $y$ is continuous, bijective, and the inverse $y^{-1}$ is itself continuous.}
\end{quotation}
For the sake of mathematical rigor we further assume that $y$ (and all other quantities) are as smooth as required, so we replace \enquote{homeomorphism} by \enquote{diffeomorphism}  ($y$ and $y^{-1}$ continuously differentiable) in the previous assumption.

A \term{referential vector} $a \in \Trm_{x} \Omega$ at a referential point $x \in \Omega$, where $\Trm_x \Omega \cong \R^d$ denotes the tangent space to $\Omega$ at $x \in \Omega$, is transformed into a \term{spatial vector} $b \in \Trm_{y(t,x)} y(t,\Omega)$ via the \term{push-forward}
\[
  b = (\di y(t,x))_* a := \nabla y(t,x) a,
  \qquad\text{where}\qquad
  \nabla y(t,x) = \left[ \frac{\partial y^j}{\partial x^k}(t,x) \right]^j_k.
\]
Note that here and in the following we switch between abstract vectors/co-vectors in the differential geometric sense and natural coordinates without further comment. Let us also adopt a simplifying \emph{notational convention}: We leave out the arguments $t$ and $x$ to any quantity such as $y$ whenever this does not cause any confusion. So, we write $y$ instead of $y(t,x)$; the meaning should be clear from the context.
 
From physical reasoning we further require that no non-zero vector is mapped to the zero vector under the push-forward $\nabla y$ and that orientations are preserved. Hence, we require
\[
  \nabla y \in \GL^+(d),
  \qquad\text{i.e.,}\qquad
  \det\, \nabla y > 0.
\]

Fundamental to all geometrically nonlinear modeling in finite-strain elasto-plasticity is the multiplicative \term{Kr\"{o}ner--Lee decomposition}~\cite{Kroner60,LeeLiu67FSEP,Lee69EPDF,GreenNaghdi71,CaseyNaghdi80,GurtinFriedAnand10,ReinaConti14}
\[
  \nabla y = E P,
\]
Here, $E$ and $P$ correspond to the \term{elastic distortion} and \term{plastic distortion}, respectively. The basic underlying assumption is that the plastic distortion is given as the \enquote{base} and the elastic distortion happens \enquote{on top of} the plastic distortion.

In general,
\[
  \curl E \neq 0
  \qquad\text{and}\qquad
  \curl P \neq 0,
\]
whence we can only speak of $E, P$ as \enquote{distortions} (this terminology is from~\cite{GurtinFriedAnand10}) and not deformation gradients (there are no corresponding $y^e, y^p$). We think of this in the following way: If $P$ is piecewise constant, everywhere except for the boundaries between different constancy domains, it is a \emph{local} gradient of a piecewise-affine plastic deformation. However, the different affine pieces do not have to fit together without creating holes or overlaps. Since \emph{globally} we assume that we end up with a continuum of mass again, these deformations have to be \emph{elastically corrected}. Consequently, plastically distorted bodies sometimes cannot fully elastically unload: If $\curl P \neq 0$, then $E = I$ is impossible since otherwise we would have $\curl \nabla y = \curl (E P) \neq 0$, which is a contradiction.


In a more abstract fashion, we can consider there to be a \emph{vector bundle} (see Chapter~10 of~\cite{Lee13} for a formal definition) over the manifold $\Omega \times [0,+\infty)$, which we denote by $\Srm \Omega = (\Srm_{(t,x)} \Omega)_{t>0, x \in \Omega}$ and call the \term{structural space}, such that at every time $t > 0$ and every point $x \in \Omega$ the associated vector space $\Srm_{(t,x)} \Omega := \R^d$ contains the \term{structural vectors}. Then, the pushforward $(\di y(t,x))_* \colon \Trm_x \Omega \to \Trm_{y(t,x)} y(\Omega)$ splits into a plastic distortion $P \colon \Trm_x \Omega \to \Srm_{(t,x)} \Omega$ and an elastic distortion $E \colon \Srm_{(t,x)} \Omega \to \Trm_{y(t,x)} y(t,\Omega)$. The maps $P, E$ act on vectors in natural coordinates as multiplication with the respective matrices, also denoted by $P, E$. Sometimes, the engineering literature refers to the \enquote{structural (intermediate) configuration} between plastic and elastic distortions. This, however, can only be understood in the sense of transformation of \emph{vectors} as before, but it is meaningless to transform points $x \in \Omega$ as there are no \enquote{structural points} (see Section~\ref{ssc:ref_struct} for more on this).

In our model, which is based on an explicit flow for the plastic part, $E$ only occurs as a \emph{derived} quantity,
\[
  E = \nabla y P^{-1}.
\]
Therefore, the issue of which invariances should be required of $E, P$, which is much discussed in the literature~\cite{GreenNaghdi71,Rice71,Mandel73,NematNasser79,Dafalias87,LubardaLee81,Naghdi90,Zbib93,CaseyNaghdi80,Mielke03a}, is not relevant here.

\subsection{Local state spaces} \label{ssc:state_spaces}

The \emph{local} state space contains all possible current states of the system at a given material point $x \in \Omega$. For the plastic distortion, we assume the existence of a real matrix Lie group $\Pfrak \subset \GL^+(d)$ (that is, a group of matrices under matrix multiplication that is also a manifold such that multiplication and inversion are smooth operations), which we call the \term{plastic distortion group}, and require
\[
  P \in \Pfrak.
\]
Thus, the full signatures of $E, P$ are $E \colon [0,T) \times \Omega \to \GL^+(d)$ and $P \colon [0,T) \times \Omega \to \Pfrak$, respectively. While the introduction of Lie group theory introduces some additional notation, it greatly clarifies some concepts and allows to consistently describe the signatures of many quantities occurring in the model. 

In metals, plastic distortions are caused by the glide of microscopic dislocations, i.e.\ defects in the crystal structure of the material, see~\cite{Kroner01,ContiTheil05,HochrainerZaiserGumbsch07,AbbaschianReedHill09,SandfeldEtAl11,HochrainerEtAl14} for some recent works in this direction. these are volume-preserving and so one often postulates \term{plastic incompressibility}
\[
  \det\, P = 1.
\]
In this case, $\Pfrak \subset \SL(d)$.

Even the assumption of plastic incompressibility, however, still does not alleviate the so-called \enquote{uniqueness problem}: The Kr\"{o}ner--Lee decomposition $\nabla y = E P$ is far from unique except in very simple cases. We adopt the usual approach to treat $P$ as an internal variable and $E = \nabla y P^{-1}$ as a derived quantity.

We denote the Lie algebra associated with the Lie group $\Pfrak$ by $\pfrak = \Lie(\Pfrak)$. Intuitively, elements of $\pfrak$ represent \enquote{infinitesimal} referential plastic distortion, or, more precisely, \term{referential plastic rates}, i.e.\ the speed by which a quantity changes. Formally, $\pfrak = \Lie(\Pfrak)$ can be defined as the tangent (vector) space to $\Pfrak$ at the identity matrix, $\pfrak = \Trm_I \Pfrak$. The fact that we consider the tangent space at the identity reflects the referential character of elements in $\pfrak$.

\begin{example}
The following are some relevant Lie groups and Lie algebras:
\begin{itemize}
  \item The Lie group $\Pfrak = \SL(d) = \set{ A \in \R^{d \times d} }{ \det\, A = 1 }$, which has the Lie algebra $\pfrak = \mathfrak{sl}(d) = \set{ A \in \R^{d \times d} }{ \tr\, A = 0 }$, is the standard choice.
  \item If we choose $\Pfrak = \SO(d) = \set{ A \in \R^{d \times d} }{ A^T = A^{-1}, \, \det\, A = 1 }$ and $\pfrak = \mathfrak{so}(d) = \set{ A \in \R^{d \times d} }{ A^T = -A }$, then only orthogonal transformations are allowed, plastic shears are forbidden.
  \item A single slip system with slip direction $s \in \R^d$, $\abs{s} = 1$, and slip plane normal $n \in \R^d$, $\abs{n} = 1$ such that $s \perp n$ can be modeled using $\Pfrak = \set{ I + \alpha (s \otimes n) }{ \alpha \in \R }$ and $\pfrak = \set{ \alpha (s \otimes n) }{ \alpha \in \R }$.
\end{itemize}
\end{example}

Motivated by the microscopic discussion about the plastic distortion equation, see~\eqref{eq:Fp_flow}, we define the \term{(referential) plastic drift} $\frac{\DD}{\DD t} P \colon [0,T) \times \Omega \to \pfrak$ (see below for the fact that $\frac{\DD}{\DD t} P \in \pfrak$) as
\[
  \frac{\DD}{\DD t} P := P^{-1} \dot{P}.
\]
To translate a spatial plastic distortion rate $\dot{P} \in \Trm_{P} \Pfrak$ relative to the (plastically) deformed configuration to the referential frame, we need to pull it back to $\pfrak \cong \Trm_I \Pfrak$. This is accomplished via a pre-multiplication of $\dot{P}$ with $P^{-1}$, yielding the referential plastic drift $\frac{\DD}{\DD t} P$. Thus, $\frac{\DD}{\DD t} P$ takes the role of $D$ in~\eqref{eq:Fp_flow}.

More abstractly, the elements of $\Lie(\Pfrak)$ can also be considered to be the left-invariant vector fields $X$ on $\Pfrak$, that is $(\di L_M)_* X = X$, where $L_M A := MA$ is the left multiplication and $(dL_M)_*$ its pushforward. Then we can express the pullback to the referential frame through $(\di L_{P^{-1}})_* \dot{P} = P^{-1} \dot{P} \in \pfrak$. Thus, the \term{referential time derivative} $\frac{\DD}{\DD t}$ acts on a curve $t \mapsto P(t) \in \Pfrak$ as
\begin{equation} \label{eq:DDt}
  \frac{\DD}{\DD t} P(t) := (\di L_{P(t)^{-1}})_* \frac{\di}{\di t} P(t)
  \quad \in \pfrak.
\end{equation}
Note that taking the referential time derivative is not a linear operation, but it obeys the usual chain rule under reparametrizations.

The cotangent space $\Trm^*_{P} \Pfrak$ to $\Pfrak$ at $P$ contains \term{stresses} (forces per unit area), cf.~\cite{Mielke03a}. An element $\sigma \in \Trm^*_{P} \Pfrak$ can be pulled back to the cotangent space $\Trm^*_{I} \Pfrak \cong \pfrak^*$ at the identity, where $\pfrak^*$ is the dual space to the Lie algebra $\pfrak$ (which is a Lie co-algebra, but we do not need this). This is accomplished via the map $\sigma \mapsto P^T \sigma \in \Trm^*_{I} \Pfrak$, which can be checked by observing that
\[
  \dot{P} : \sigma = (P^{-1} \dot{P}) : (P^T \sigma),
\]
where $A : B := \sum_{i,j} A^i_j B^i_j$ is the \emph{Frobenius product} of the matrices $A,B \in \R^{d \times d}$, which here acts as the duality pairing between $\pfrak$ and $\pfrak^*$. The elements of $\pfrak^*$ therefore are \term{referential stresses}.

Furthermore, we assume that the material body possesses \term{internal variables} $z \in \Zfrak$, where $\Zfrak = \R^m$ is the canonical vector Lie group describing the internal variables of the system (for example, dislocation density in metals) with associated \emph{abelian} (i.e.\ trivial) Lie algebra $\zfrak = \Lie(\Zfrak) \cong \R^m$. General Lie groups $\Zfrak$ are also possible, but a simple vector will suffice for us here. According to~\eqref{eq:DDt}, in this commutative Lie group we have for a curve $t \mapsto z(t) \in z(t) \in \Zfrak$ that
\[
  \frac{\DD}{\DD t} z(t) = \dot{z}(t).
\]
since left \enquote{multiplication} is simply addition and hence $(\di L_{z(t)^{-1}})_* = \id$.

We bundle together the plastic distortion and internal variables (at a point) and define the \term{(local) full internal state space} as the product Lie group
\[
  \Hfrak := \Pfrak \times \Zfrak,
\]
which has Lie algebra $\hfrak = \Lie(\Hfrak) \cong \pfrak \times \zfrak$, where $\pfrak = \Lie(\Pfrak)$ and $\zfrak = \Lie(\Zfrak)$. In this way we consider the plastic distortion to be part of the internal state of the material, using the notation
\[
  H := (P,z) \in \Hfrak.
\]

\section{Energy and dissipation} \label{sc:energy}

In this section, we describe the energetics and basic dynamics of the two \emph{fundamental motions} that we consider in our modelling approach: \emph{elastic minimization}, for which $\dot{P}, \dot{z} = 0$, and \emph{elasto-plastic relaxation}, for which $\dot{y} = 0$. The next section will combine these basic motions into a full dynamic model, keeping in mind the relative speed by which these two motions progress.

\subsection{Elastic free energy}

The energetics of the material are governed by the \term{free energy functional} 
\[
  \Wcal[y,P,z] := \int_\Omega W(x,\nabla y(t,x),P(t,x),z(t,x)) \dd x,
\]
where $W \colon \Omega \times \GL^+(d) \times \Hfrak \to \R$ is the \term{energy density} (recall $\Hfrak = \Pfrak \times \Zfrak$), which associates to each local state an energy per referential unit volume. We further split $W$ as
\[
  W(x,F,P,z) = W_e(x,FP^{-1}) + W_h(x,z).
\]
Here, $W_e \colon \Omega \times \GL^+(d) \to \R$ is the \term{elastic energy density} and $W_h \colon \Omega \times \Zfrak \to \R$ is the \term{hardening energy density}.

We assume $W_e,W_h$ to be as smooth as necessary and $W_e$ to satisfy the \term{objectivity (frame-indifference)}
\begin{equation} \label{eq:frame_indiff}
  W_e(x,RE) = W_e(x,E)  \qquad\text{for all $x \in \Omega$, $R \in \SO(d)$, $E \in \GL^+(d)$.}
\end{equation}

%

Often, $W_e,W_h$ also satisfy certain material symmetries, such as $W_e(x,ES) = W_e(x,E)$ for all $S$ from a \term{symmetry group} $\Sfrak \subset \GL^+(d)$. The most common such material symmetry is \term{isotropy}, for which $\Sfrak = \SO(d)$, the group of orthogonal $(d \times d)$-matrices with positive determinant, see Section~3.4 in~\cite{Ciarlet88}; a recent study of the isotropic case is in~\cite{GrandiStefanelli15}.

As already mentioned before, if $\curl P \neq 0$, no complete elastic unloading is possible, as $E \equiv R \in \SO(d)$ would then contradict $\curl E P = 0$. Thus, not all of the free energy might be \enquote{free} to do work. However, as this is essentially a consequence of the requirement of geometric compatibility, we stick to the usual terminology. 

With an external \term{bulk loading} $f(t,x) \in \R^d$ we also define the \term{total energy functional}
\begin{align}
  \Ecal[t,y,P,z] &:= \Wcal[y,P,z] - \int_\Omega f(t,x) \cdot y(t,x) \dd x  \notag\\
  &\phantom{:}= \int_\Omega W(x,\nabla y(t,x),P(t,x),z(t,x)) - f(t,x) \cdot y(t,x) \dd x.  \label{eq:Ecal}
\end{align}

\subsection{The virtual power principle for purely elastic movements}

Let $\Omega' \subset \Omega$ be a referential subdomain of the body and denote by $n$ the unit outward normal on $\partial \Omega'$. For the first fundamental motion of the system, namely the purely elastic motion, we assume $\dot{P}=0$, $\dot{z}=0$, which together with $\nabla y = E P$ implies
\[
  \dot{E}=\nabla\dot{y}P^{-1}.
\]
Here and in the following we leave out the arguments $(t,x)$ to make the formulas more readable.

The \term{internal power} expended \emph{within} $\Omega'$ can be computed as
\begin{align*}
  \Ical(\Omega') &= \int_{\Omega'} \frac{\di}{\di t} W(x,\nabla y,P,z) \dd x \\
           &= \int_{\Omega'}  
               \DD_F W(x,\nabla y,P,z) : \nabla \dot{y}  \dd x \\
           &= \int_{\partial \Omega'} \DD_F W(x,\nabla y,P,z) n \cdot \dot{y} \dd a
              + \int_{\Omega'} -\Diverg [\DD_F W(x,\nabla y,P,z) ] \cdot \dot{y} \dd x.
\end{align*}
Notice that $\DD_F W(x,F,P,z) \in \Trm^*_F \GL^+(d)$, which is the cotangent space to $\GL^+(d)$ at $F$; then, the Frobenius product \enquote{$:$} can also be interpreted as the application of this covector on the vector $\nabla \dot{y}$.

On the other hand, the \term{external power} expended \emph{on} $\Omega'$ is given by 
\[
  \Pcal(\Omega') = \int_{\partial \Omega'} t(n) \cdot \dot{y} \dd a
  + \int_{\Omega'} f(t) \cdot \dot{y} \dd x
  - \int_{\Omega'} \rho \ddot{y} \cdot \dot{y} \dd x,
\]
where $t(n) \in \R^d$ is the surface traction in direction $n$, $f(t) = f(t,x) \in \R^d$ is the external bulk loading and $\rho > 0$ is the mass density. More complex external forces are of course possible as well, but we try to keep the exposition simple. Then, we assume the fundamental \emph{Principle of Virtual Power} (see Chapter~92 in~\cite{GurtinFriedAnand10}):
\begin{quotation}
  \textit{The internal and external powers are equal, $\Ical(\Omega') = \Pcal(\Omega')$, for all allowed choices of $(F,P,z)$ and all $\Omega' \subset \Omega$.}
\end{quotation}
In this context, $y$ and $E$ are referred to as \emph{virtual} because they might not be attained in a given evolution (but they are \emph{attainable}). Thus, with our $\Ical(\Omega')$ and $\Pcal(\Omega')$, we get
\begin{align*}
  &\int_{\partial \Omega'} \bigl( \DD_F W(x,\nabla y,P,z) n - t(n) \bigr) \cdot \dot{y} \dd a
    + \int_{\Omega'} \bigl( \rho \ddot{y} - \Diverg [ \DD_F W(x,\nabla y,P,z) ] - f \bigr) \cdot \dot{y} \dd x \\
  &\qquad = 0.
\end{align*}
Then, since $\Omega' \subset \Omega$ was arbitrary, we get the \term{power balance} (we drop the boundary equation since we do not need this information in the following)
\[
  \bigl( \rho \ddot{y} - \Diverg [\DD_F W(x,\nabla y,P,z)] \bigr) \cdot \dot{y}
       = f \cdot \dot{y}.
\]
The virtual rate $\dot{y}$ can attain any value, and so we conclude
\begin{equation} \label{eq:elastic_force_balance_from_power}
  \rho \ddot{y} - \Diverg [\DD_F W(x,\nabla y,P,z)] = f.
\end{equation}

\begin{remark}
If we instead carry out the preceding computation for $W_e$, we get
\[
  \rho \ddot{y} - \Diverg [ \DD_E W_e(x,E) P^{-T} ] = f.
\]
However, it can be easily checked that $\DD_E W_e(x,E) P^{-T} = \DD_F W(x,\nabla y,P,z)$, and so this is the same as~\eqref{eq:elastic_force_balance_from_power}.
\end{remark}


Besides the power balance expressed in the \emph{Principle of Virtual Power}, we can also consider the balance of physical forces as fundamental. This gives rise to an alternative derivation of~\eqref{eq:elastic_force_balance_from_power}: The \term{linear momentum} of a (referential) subset $\Omega' \subset \Omega$ is
\[
  \int_{\Omega'} \rho \dot{y} \dd x,
\]
where $\rho > 0$ is the density of the material.

By \emph{Newton's Second Law}, the change of linear momentum is equal to all forces acting on $\Omega'$. There are two such forces:
\begin{enumerate}
  \item The body force $\displaystyle \int_{\Omega'} f \dd x$.
  \item The surface traction $\displaystyle \int_{\partial \Omega'} t(n) \dd a = \int_{\Omega'} \Diverg T_R \dd x = \int_{\Omega'} \Diverg [\DD_F W(x,\nabla y,P,z)] \dd x$.
\end{enumerate}
Thus, we have
\[
  \int_{\Omega'} \rho \ddot{y} \dd x
  = \frac{\di}{\di t} \int_D \rho \dot{y} \dd x
  = \int_{\Omega'} f + \Diverg [\DD_F W(x,\nabla y,P,z)] \dd x
\]
As $\Omega' \subset \Omega$ was arbitrary, this is equivalent to our previous force balance~\eqref{eq:elastic_force_balance_from_power}.

From now on we also assume (one can also derive a more complex theory without this):
\begin{quotation}
  \textit{Inertial forces can be neglected.}
\end{quotation}
Then, we get the \term{equilibrium equation}
\[
  - \Diverg [ \DD_F W(x,\nabla y,P,z) ] = f.
\]


\subsection{The virtual power principle for elasto-plastic relaxation movement} \label{ssc:relax_vpower}

Now assume $\dot{y}=0$, whence on each subdomain $\Omega' \subset \Omega$ the external power expended on $\Omega'$ is zero. On the other hand, for the internal power we obtain
\[
\Ical(\Omega')= \int_{\Omega'} \DD_{P} W(x,\nabla y,P,z) : \dot{P} + \DD_z W(x,\nabla y,P,z) \cdot \dot{z} + X_p : \dot{P} + X_z \cdot \dot{z} \dd x,
\]
where
\begin{enumerate}
\item $X_p \colon \Omega \to T^*_{P}\Pfrak$ is the generalized stress power-conjugate to $\dot{P}$ (recall that $T^*_{P}\Pfrak$ is the cotangent space to $\Pfrak$ at $P$, that is, the dual space to the tangent space $\Trm_{P} \Pfrak$), and
\item $X_z \colon \Omega \to T^*_z\Zfrak$ is the generalized stress power-conjugate to $\dot{z}$.
\end{enumerate}
The two generalized stresses $X_p, X_z$ are the stresses that the system has to work against to deform plastically.

Since, again by the \emph{Principle of Virtual Power}, $\Ical(\Omega') = \Pcal(\Omega') = 0$ for any subdomain $\Omega' \subset \Omega$, we obtain the local balance of powers
\begin{align} \label{eq:plastic_balance}
\DD_{P} W(x,\nabla y,P,z) : \dot{P} + \DD_z W(x,\nabla y,P,z) \cdot \dot{z} = -X_p : \dot{P} - X_z \cdot \dot{z}.
\end{align}
As $\dot{P}$ and $\dot{z}$ can take any value, we conclude the \term{elasto-plastic relaxation balance}
\[
  \left\{\begin{aligned}
    -\DD_{P} W(x,\nabla y,P,z) &= X_p \\
    -\DD_z W(x,\nabla y,P,z)   &= X_z.
  \end{aligned}\right.
\]
Upon defining the \term{referential plastic stress} $T_p \colon \Omega \to \pfrak^*$ as
\[
  T_p := - P^T \DD_{P} W(x,\nabla y,P,z)
\]
we can rewrite the elasto-plastic relaxation balance as
\begin{equation} \label{eq:elasto-plastic_relax_balance}
  \left\{\begin{aligned}
    T_p               &= P^T X_p \\
    -\DD_z W(x,\nabla y,P,z)   &= X_z.
  \end{aligned}\right.
\end{equation}
Since $\DD_{P} W(x,\nabla y,P,z) \in \Trm^*_{P} \Pfrak$, we indeed have $T_p(F,P,z) \in \pfrak^*$, see Section~\ref{ssc:state_spaces}.

Let us also indicate how this rule can be written for $W_e,W_h$ and the corresponding elastic stress: $T_p = - P^T \DD_{P} W(x,\nabla y,P,z)$ lies in $\pfrak^*$, the dual space to $\pfrak$. In coordinates, we represent $\pfrak,\pfrak^*$ by matrices and the duality product by the Frobenius product. Of course, $\pfrak,\pfrak^*$ are in general strict subspaces of $\R^{d \times d}$, so there are additional conditions that have to be satisfied by matrices in these spaces. During coordinate calculations, these side constraints can become \enquote{lost} and so at the end we need to project onto the corresponding spaces again.

We have for any $D \in \pfrak$, and using $E := \nabla y P^{-1}$,
\begin{align*}
  T_p : D &= - P^T \DD_{P} W(x,\nabla y,P,z) : D \\
  &= - \DD_{P} W_e(x,E) : (P D)  \\
  &= \DD_E W_e(x,E) : [\nabla y D P^{-1}] \\
  &= \nabla y^T \DD_E W_e(x,E) P^{-T} : D,
\end{align*}
since for the derivative of $P \mapsto \nabla y P^{-1}$ in direction $PD \in \Trm_P \Pfrak$, we have
\[
  \frac{\di (\nabla y P^{-1})}{\di P} [PD]
  = - \nabla y P^{-1} PD P^{-1}
  = - \nabla y D P^{-1}
\]
by the well-known formula $\partial_t ([A+tQ]^{-1}) = - A^{-1} Q A^{-1}$. Thus,
\begin{equation} \label{eq:ps_proj}
  T_p = \mathrm{proj}_{\pfrak^*} \bigl[ \nabla y^T \DD_E W_e(x,E) P^{-T} \bigr],
\end{equation}
where we denote the orthogonal projection onto $\pfrak^*$ by \enquote{$\mathrm{proj}_{\pfrak^*}$} (for example, if $\Pfrak = \SL(d)$, $\pfrak = \mathfrak{sl}(d)$, then also $\pfrak^* \cong \mathfrak{sl}(d)$ and $\mathrm{proj}_{\pfrak^*}(U) = \dev U = U - (\tr U)I/d$).

We define the \term{total generalized stress} $\Sigma \colon \Omega \to \hfrak^*$ (recall that $\hfrak$ is the Lie algebra to $\Hfrak$ and $\hfrak^*$ is its dual) through
\begin{equation}\label{eq:Sigma}
  \Sigma = \Sigma(x,F,P,z) := \bigl( T_p(x,F,P,z) , -\DD_z W(x,\nabla y,P,z) \bigr),
\end{equation}
our balance~\eqref{eq:elasto-plastic_relax_balance} is then concisely expressed as
\begin{equation} \label{eq:preflow}
  \Sigma = ( P^T X_p, X_z).
\end{equation}

Finally, we remark that from the frame-indifference of $W$ by assumption~\eqref{eq:frame_indiff}, one may conclude that also $T_p$ is frame-indifferent, i.e.\
\[
  T_p(RF,P,z) = T_p(F,P,z)  \qquad\text{for all $R \in \SO(d)$.}
\]

It is now straightforward to implement thermodynamic principles for the elasto-plastic relaxation phase by appealing to the \term{free-energy imbalance} as stated e.g.\ in Section 27.3 of~\cite{GurtinFriedAnand10}, which is itself a consequence of the \emph{Second Law of Thermodynamics}. The free-energy imbalance means that in each subregion $\Omega' \subset \Omega$,
\[
  \frac{\dd}{\dd t}\Wcal(\Omega')-\Pcal(\Omega')=-\Delta(\Omega') \leq 0,
\]
with the nonnegative dissipation $\Delta(\Omega')$. By the \emph{Principle of Virtual Power} we also have $\Pcal(\Omega')=\Ical(\Omega')$. Thus we obtain
\[
  \frac{\dd}{\dd t}\Wcal(\Omega')-\Ical(\Omega')=-\Delta(\Omega') \leq 0.
\]
Since we are considering the elasto-plastic relaxation phase, $\nabla \dot{y} = 0$. Then, use
\[
  \frac{\dd}{\dd t}W(x,\nabla y,P,z) = \DD_{P} W(x,\nabla y,P,z) : \dot{P} +\DD_z W(x,\nabla y,P,z)\cdot\dot{z}
\]
together with~\eqref{eq:plastic_balance} and localize, which yields 
\[
  X_p : \dot{P} + X_z\cdot\dot{z}=\delta\geq0,
\]
where $\delta \colon [0,T) \times \Omega \to [0,+\infty)$ is the \term{dissipation rate}. Then, using~\eqref{eq:preflow}, we have for the component-wise scalar product $\Sigma\circ\frac{\DD}{\DD t} H$ that
\begin{equation}\label{eq:thermodyn_dissip}
  \frac{\DD}{\DD t} H \circ \Sigma = \frac{\DD}{\DD t} P : (P^T X_p) + \dot{z} \cdot X_z
  = \dot{P} : X_p + \dot{z} \cdot X_z = \delta \geq 0.
\end{equation}
Inequality~\eqref{eq:thermodyn_dissip} expresses the \term{irreversibility} of plastic flow whenever $\delta > 0$. In contrast, purely elastic deformations are of course reversible.

\subsection{Dissipation potential and flow rule}

So far we have not constitutively specified $X_p$ and $X_z$ in~\eqref{eq:preflow}. In fact, we prefer -- as is usual in much of the mathematical treatment of elasto-plasticity -- to assume that~\eqref{eq:preflow} can be written as a differential inclusion or, equivalently, as a variational inequality. This is chiefly a structural assumption, in this context see Section~4.4 in~\cite{Fremond02} as well as~\cite{ZieglerWehrli87,Silhavy97,Mielke03a}. At this point we just consider the situation \emph{in general} and do not make any further assumptions (rate-dependence or rate-independence).


So, assume we have a \term{dissipation (pseudo)potential} $R \colon \hfrak \to [0,+\infty]$ (recall that $\hfrak = \pfrak \times \zfrak$) with the properties
\begin{quotation}
  \textit{$R$ is convex, lower semicontinuous, and strictly positive.}
\end{quotation}
Here, strict positivity means that $R(\Lambda) \circ \Lambda > 0$ for $\Lambda \neq 0$ (which is equivalent to $R(\Lambda) > 0$ for $\Lambda \neq 0$ whenever $R$ is positively homogeneous of any order).

We furthermore assume that the \term{flow rule}~\eqref{eq:preflow} can be written in the form of the \term{Biot inclusion} (cf.~Section~4.4 in~\cite{Fremond02} and~\cite{Silhavy97})
\begin{equation} \label{eq:flow}
  \Sigma \in \partial R\biggl( \frac{\DD}{\DD t} H \biggr), \qquad\text{where}\qquad
  \frac{\DD}{\DD t} H := \biggl( \frac{\DD}{\DD t} P, \frac{\DD}{\DD t} z \biggr) = (D, \dot{z}) \in \hfrak.
\end{equation}
Here, $\frac{\DD}{\DD t} H$ is called the \term{total drift} and $\partial$ denotes the convex subdifferential of $R$ at $\frac{\DD}{\DD t} H$, that is, $\partial R\bigl(\frac{\DD}{\DD t} H\bigr)$ consists of all those \term{flow stresses} $\Sigma \in \hfrak^*$ that satisfy
\[
  R\biggl( \frac{\DD}{\DD t} H \biggr) + \biggl(\Lambda - \frac{\DD}{\DD t} H \biggr) \circ \Sigma \leq R(\Lambda) \qquad
  \text{for all $\Lambda \in \hfrak$,}
\]
where \enquote{$\circ$} is the component-wise scalar product between $\hfrak$ and $\hfrak^*$, that is, for $\Lambda = (A,c) \in \hfrak$ and $\Gamma = (B,d) \in \hfrak^*$ we have $\Lambda \circ \Gamma := A : B + c \cdot d$.

If we for the moment (and for the purpose of illustration only) assume that our dissipation potential $R$ is differentiable, then the flow rule~\eqref{eq:flow} expresses that
\[
  \Sigma = (T_p,-\DD_z W(x,\nabla y,P,z)) = (P^T X_p, X_z) = \DD R\biggl( \frac{\DD}{\DD t} H \biggr),
\]
which is just~\eqref{eq:preflow} in a special form. In particular, we assume convexity and lower semicontinuity of $R$, which are not present in~\eqref{eq:preflow}. If $R$ is positively $k$-homogeneous, then convexity of $R$ expresses the following intuitive constraint: When the system is moving in direction $\Lambda = (1-\theta) \Lambda_0 + \theta \Lambda_1 \neq 0$, where $\Lambda_0, \Lambda_1 \in \hfrak \setminus \{0\}$, $\theta \in (0,1)$, then the frictional power is $\DD R(\Lambda) \circ \Lambda = k R(\Lambda)$ (by Euler's positive homogeneity theorem); alternatively, we could oscillate very quickly between rates $\Lambda_0, \Lambda_1$ with time-fractions $1-\theta$ and $\theta$, respectively, which would result in the frictional power $(1-\theta) k R(\Lambda_0) + \theta k R(\Lambda_1)$. The convexity tells us that this oscillatory path expends more energy, as should intuitively be the case for \enquote{reasonable} materials. Moreover, it can be shown that the \emph{Maximum Plastic Work Principle}, which is a strengthening of the \emph{Second Law of Thermodynamics}, implies convexity of $R$, see p.~57--59 in~\cite{HanReddy13}. The lower semicontinuity is a minimal continuity assumption and can again be justified on physical grounds or out of mathematical necessity. Finally, the strict positivity means that energy is dissipated if and only if the material deforms plastically.

Note that in many applications $R$ does not depend on the state variables $(y,P,z)$ since we can often model changes in the size or shape of the elastic stability domain (defined below) through residual stresses. However, one may extend the theory to incorporate this constraint as well, but chose not to do this here for the sake of notational clarity.

The reason we specify the (primal/dual) flow rule in terms of the  drift $(D,\dot{z})$ and not in terms of the current flow velocity $(\dot{P},\dot{z})$ is that when the material flows, the flow rule needs to \enquote{flow with the material}, and therefore $\frac{\DD}{\DD t} H$ ought to be a \emph{referential} (or \emph{structural}) quantity. Also see Section~\ref{ssc:ref_struct} for more on this.

Finally, define the \term{dual dissipation potential} $R^* \colon \hfrak^* \to [0,+\infty]$ as the convex conjugate function of $R$, that is,
\[
  R^*(\Sigma) := \sup\, \setb{ \Sigma \circ \Lambda  - R(\Lambda)}{ \Lambda \in \hfrak }.
\]
Then, by standard results in convex analysis, see for example~\cite{Rock70CA}, we have that $R^*$ is also convex and lower semicontinuous. Moreover, the flow rule~\eqref{eq:flow} is equivalent to the \term{dual flow rule}, which is sometimes called the \term{Onsager inclusion},
\begin{equation} \label{eq:flow_dual}
  \frac{\DD}{\DD t} H \in \partial R^*(\Sigma).
\end{equation}
See~\cite{HanReddy13} for more on these duality aspects, albeit in a geometrically linear framework.

So far, we have not specified anything about the dependence of $R$ on $\frac{\DD}{\DD t} H$ beyond convexity and lower semicontinuity and in fact we will later use several dissipation potentials.


According to the Coleman--Noll procedure~\cite{ColemanNoll63,Gurtin2000,GurtinFriedAnand10}, thermodynamic reasoning should give \emph{constitutive restrictions}. In this spirit, combining~\eqref{eq:thermodyn_dissip} with the flow rule~\eqref{eq:flow} yields the \term{positivity} condition on the dissipation potential $R$ and its dual $R^*$:
\begin{equation} \label{eq:R_Rs_positivity}
  \frac{\DD}{\DD t} H \circ \partial R\biggl( \frac{\DD}{\DD t} H \biggr) \geq 0,  \qquad
  \partial R^*(\Sigma) \circ \Sigma \geq 0
\end{equation}
where the product \enquote{$\circ$} is understood to act on every stress in $\partial R\bigl(\frac{\DD}{\DD t} H\bigr)$ individually. This is a weaker version of the strict positivity we required above. Furthermore, it is elementary to see that for a convex function $g \colon X \to [0,+\infty]$ with $g(0) = 0$ it always holds that $\partial g(x) \cdot x \geq 0$, hence the above condition~\eqref{eq:R_Rs_positivity} is automatically satisfied for convex $R$, giving another reason to require convexity as a structural property.

Finally, for a (time-differentiable) process $u = (y,P,z)$ we define the \term{(total) dissipation} over an interval $[s,t]$ as
\[
  \Diss(u;[s,t]) := \int_s^t \delta(\tau) \dd \tau
  = \int_s^t \frac{\DD}{\DD \tau} H(\tau) \circ \Sigma(\tau) \dd \tau \geq 0.
\]
We will later see how to define the dissipation for lower-regularity processes.

\subsection{Elastic stability domain}

It is a fundamental property of elasto-plastic processes that if the generalized stress $\Sigma$ lies in the interior of an \term{elastic stability domain} $\Scal \subset \hfrak^*$, then no plastic flow takes place. The thinking here is that below the stress threshold $\partial \Scal$, which is called the \term{yield surface}, no plastic slip can be activated. This is in very good agreement with experiments, see Chapter~5 in~\cite{LemaitreChaboche90}. We assume
\[
  \text{\textit{$\Scal \subset \hfrak$ is a closed convex neighborhood of the origin.}}
\]
In a more elaborate model, we could let $\Scal$ depend on the current internal state $(P,z) \in \Hfrak$, but this introduces additional complications (for instance, when the system moves toward $\Scal(P,z)$, this set may \enquote{move away} and we may never reach it). Many interesting models, however, model changes in the elastic stability domain through a residual plastic stress and we confine ourselves to this setup.

In the usual rate-independent plasticity theory, $\Scal = \partial R_1(0)$ for the rate-independent flow potential $R_1$, and it is furthermore assumed that $\Sigma$ can never leave $\Scal$ and plastic flow is allowed only if the stress lies on the yield surface, i.e.\ $\Sigma \in \partial \Scal$. We here diverge from this usual modeling and allow $\Sigma$ to be outside $\Scal$, at least for a very short time. This occurs if the elastic minimization selects a state $(F,P,z)$ with $\Sigma(F,P,z) \notin \Scal$. This state, however, is \emph{transient} in the sense that the material will try to correct the situation by following an elasto-plastic relaxation path, which at the terminal point will restore $\Sigma \in \Scal$. As we allow this relaxation to occur with infinite speed relative to the global (slow) time, this relaxation action \emph{is not visible} on the slow (global) timescale. However, if the system jumps to a far away state, then, on a \enquote{fast} time scale, the relaxation path is not contracted and should form part of the solution concept of our model.


\begin{remark}
It should be noted that the projection in~\eqref{eq:ps_proj} is of great importance for the consistency of the theory developed here: It ensures that only the \emph{plastic} part of the stress remains in $T_p$. This is important, because only this stress can possibly be corrected by plastic flow. To illustrate this, consider a state $P = I$, $E = \alpha I$, $\alpha \neq 1$ for $\Pfrak = \SL(d)$ (and no internal variables). The material cannot decrease its elastic energy, say $W_e(x,E) := \frac{1}{2}\abs{E}^2 = \alpha^2/2$, through an elasto-plastic relaxation flow, since a rate in $\pfrak = \mathfrak{sl}(d)$ cannot change the determinant of $E$. Thus, in order to always end up with $\Sigma \in \Scal$ after the elasto-plastic relaxation, it needs to hold that for $E = \alpha I$, $\alpha \neq 1$ we have $T_p = 0$. Indeed, while $\DD_E W_e(x,E) = \alpha I$ is \emph{not} zero, it still holds that $T_p = 0$ thanks to the projection onto deviatoric matrices in~\eqref{eq:Sigma}.
\end{remark}

\begin{example} \label{ex:D_nonconvex}
The \term{elastic deformation domain} $\Dcal(P,z) \subset \GL^+(d)$ is the preimage of the map taking the total deformation $F$ to a generalized stress $\Sigma$ while holding $P,z$ constant: 
\[
  \Dcal(x,P,z) := \setb{ F \in \GL^+(d) }{ \Sigma(x,F,P,z) \in \Scal }.
\]
It models the set of deformation gradients $\nabla y$ that can be attained by a stable state. Even when the elastic energy density $W_e$ is convex, the geometry of the elastic deformation domain can be very complex. In particular, it does not follow that the elastic deformation domain $\Dcal$ is convex. Indeed, consider the following example (without internal variables): Let for an $0 < \eps < 1/\sqrt{2}$ the elastic stability domain be
\[
  \Scal = \setb{G\in \Rbb^{2\times2}}{ \text{$\abs{G - \diag(x,x)} \leq \eps$,\, $x\in[0,1]$} },
\]
which is closed, convex, and a neighborhood of $0$. Also let $W_e(x,E)=\frac{1}{2}\abs{E}^2$. Then, according to~\eqref{eq:elasto-plastic_relax_balance},
\[
  \Sigma(F,P) = F^T \DD_E W_e(x,E) P^{-T}
  = F^T E P^{-T} = (F^T F)[P^T P]^{-1}.
\]
At $P = I$, the elastic deformation domain is
\[
  \Dcal(I) = \setb{ F\in \Rbb^{2\times2} }{ \Sigma(F,I) \in \Scal }
  = \setb{ F\in \Rbb^{2\times2} }{ F^T F \in \Scal },
\]
which is not convex: the matrices $\diag(-1,1)$ and $\diag(1,1)$ are in $\Dcal(I)$, but their average $\diag(0,1)$ is not, as a simple calculation shows.
\end{example}

\subsection{Hardening}

In its most general form, \term{hardening} describes the process by which the \emph{effective} elastic stability domain changes, in particular expands or contracts, due to a change in the internal variables. Hardening is usually anisotropic and is due to a variety of microscopic effects like dislocation entanglement and generation.

In \term{isotropic hardening}, the elastic stability domain $\Scal$ remains centered around the origin but can expand in what is called \term{positive hardening} and contract in \term{negative hardening} or \term{softening}. In \term{kinematic hardening}, the elastic stability domain is translated (usually in direction of the plastic flow). Combined, these two effects give a first approximation to the often-observed phenomenon that an increase in tensile yield strength goes along with a decrease in compressive yield strength, called the \term{Bauschinger effect}, which, however, in general is more complex, see for instance Section~3.3.7 in~\cite{LemaitreChaboche90} and \cite{KassnerEtAl09}.


\begin{example} \label{ex:vonMises_istrop_kinemat}
In the often-considered \term{Mises isotropic--kinematic hardening}, the elastic stability domain depends on two internal variables, $S \in \pfrak^* \subset \R^{d \times d}$ and $\eta \in \R$, so that $\Zfrak = \pfrak^* \times \R$ (considered as a vector in $\R^{d^2+1}$). Then, the elastic stability domain is
\[
  \Scal = \setB{ \Gamma=(T_p,S,\eta) \in \hfrak^* }{ \abs{\dev(T_p-S)} + \eta - \textstyle\frac{2}{3} \sigma_0 \leq 0 },
\]
where $\sigma_0 > 0$ is a constant (the initial tensile yield strength) and $\abs{A} = \abs{A}_F = [\tr (A^T A)]^{1/2}$ is the Frobenius norm. The internal variables split into $S \in \pfrak^*$ and $\eta \in \R$, called \term{backstresses}. This means that the elastic stability domain is translated with $S$ and dilated with $\eta$ (a scalar stress). For the hardening energy $W_h$ we set (for instance)
\[
  W_h(S,\eta) := \frac{1}{2}g(\abs{S}^2) + h(\eta),
\]
with $g,h \colon \R \to \R$ such that $g(0) = h(0) = 0$. Notice that our $\Scal$ above does not depend on the internal variables $S,\eta$, which is characteristic for the approach using backstresses. In fact, a dependence of $\Scal$ on the internal variables seems to be needed only rarely. In purely \term{isotropic hardening} the internal variable $S$ is not present. Using a different matrix norm, one can get different shapes of the yield surface.
\end{example}



\subsection{Referential versus structural formulation} \label{ssc:ref_struct}

There is some disagreement in the literature over what should be called the \enquote{undistorted} crystal lattice, see for example Section~91.2 in~\cite{GurtinFriedAnand10} and~\cite{Naghdi90}. Of course, the \emph{observed} lattice (which can be visualized for example through orientation-imaging microscopy) is the one that is already plastically distorted from the original \emph{referential} lattice, so it is located in the \emph{structural} space $\Srm \Omega$ defined in Section~\ref{ssc:state_spaces}. We can always transform between referential and structural vectors by applying $P$, which gives the structural vector $P m$ corresponding to a referential vector $m$, and the corresponding inverse transformation. 



In mathematical terms, we need to decide whether we should formulate the flow rule and all other plastic quantities with respect to the reference frame or with respect to the structural frame, where the latter appears to be more popular~\cite{Mielke03a,GurtinFriedAnand10}. All quantities have versions in both frames. Indeed, in the structural frame, the plastic rate is measured via the \term{structural plastic distortion rate}
\[
  L := \dot{P} P^{-1} = P \biggl(\frac{\DD}{\DD t} P\biggr) P^{-1},
\]
which is a tensor that maps structural vectors to structural vectors since $P$ maps material vectors to structural vectors. The \enquote{structural stress} is the \term{Mandel stress}
\[
  M_e := P^{-T} T_p P^T
  = - \DD_{P} W(x,\nabla y,P,z) P^T.
\]
It is easy to see that $L$ and $M_e$ are again conjugate in the sense that $M_e : L$ is equal to the dissipation $\delta$, in complete analogy to the fact that $T_p$ and $D$ are conjugate. Furthermore, it turns out that the referential and the structural flow rule are equivalent if $R$ is allowed to depend on $P$. Indeed, a computation shows that~\eqref{eq:flow} is equivalent to
\[
  (M_e, -\DD_z W(x,\nabla y,P,z)) \in \partial_{(L,\dot{z})}\tilde{R}(P,z,L,\dot{z}),
\]
where
\[
  \tilde{R}(P,z,G,\zeta):= R(P^{-1} G P,\zeta).
\]

Considering the abstract algebraic structure, the transition from referential to structural space is accomplished via \emph{conjugation}: Define the \term{adjoint mapping} $\Ad_Q \colon \pfrak \to \pfrak$ and the \term{dual adjoint mapping} $\Ad^*_Q \colon \pfrak^* \to \pfrak^*$ through
\[
  \Ad_Q(A) := QAQ^{-1},  \qquad
  \Ad^*_Q(A) := Q^{-T}AQ^T.
\]
Then,
\[
  L = \Ad_P \biggl( \frac{\DD}{\DD t} P \biggr),  \qquad
  M_e = \Ad^*_{P}(T_p).
\]
Since $\Ad_{P}$ and $\Ad^*_{P}$ are isomorphisms, the formulations are indeed completely interchangeable. In the isotropic case, also $R = \tilde{R}$, see~\cite{GrandiStefanelli15}.

We remark that while the referential and the structural formulation are in principle equivalent, there may be good reasons to choose one over the other, at least in the non-isotropic case. It is quite possible that one of the $R, \tilde{R}$ is independent of $P$, whereas the other is not. In fact, if the anisotropy originates in the crystal lattice (for example different behavior along the crystal vectors in crystal plasticity), then the structural $\tilde{R}$ has a good chance of being independent of $P$ (and hence \enquote{simpler}). In this case, the whole model should be formulated in the structural frame, which is an easy translation. On the other hand, if the anisotropy is due to mesoscale laminations, then $R$ may be independent of $P$.

In the absence of the reasons just alluded to, we however believe that there is at least an \enquote{aesthetic} reason to prefer formulating the model in the reference frame: Since $P$ is in general not a gradient, there are no \enquote{structural points}, only material points. The engineering literature sometimes considers points $x_S$ in an imaginary \enquote{structural space}, but this is not a rigorous notion. Moreover, if we do not require the plastic incompressibility $\det P = 1$ (soils, for example, are plastically compressible), which in any case should be treated as just a \emph{constitutive} assumption, then the Mandel stress should reflect this and we really ought to use
\[
  M_e^S(x_S) := J_p(x_R) P(x_R)^{-T} T_p(x_R) P(x_R)^T
  \qquad\text{with}\qquad
  J_p(x_R) := \det P(x_R).
\]
In other words, the points $x_S$ from the structural space are just points in $\Omega$, but the \emph{structural manifold} $\Omega$ is now equipped with a different metric. All of this seems rather cumbersome and in light of the equivalence exposed above and also the microscopic motivation of plastic flow in~\eqref{eq:Fp_flow}, we prefer to work in the referential frame.


\section{Global time evolution} \label{sc:evolution}

In reality, plastic flow is \emph{rate-dependent} (viscous), but only slightly so below absolute temperatures of approximately $0.35 \vartheta_m$, where $\vartheta_m$ is the melting temperature of the material, see~Section~78 in~\cite{GurtinFriedAnand10}. For instance, in a commonly-used power viscosity law (see Section~5.4 in~\cite{LemaitreChaboche90}) the stress depends on the rate with exponent $1/N$, that is, $\DD R(\Lambda) \sim \abs{\Lambda}^{1/N}$. For example, steel with 35\% carbon at 450 \textdegree{}C has $N = 15$ and the titanium-aluminium alloy TA6V at 350 \textdegree{}C has $N = 120$; more values can be found in Table~6.2 of~\cite{LemaitreChaboche90}. The movement of the system is directed in such a way as to move the stress $\Sigma$ toward $\partial \Scal$, where the movement stops as the internal friction stress threshold is no longer exceeded. It can be seen that the larger $N$ is, the faster this \enquote{relaxation} takes place. Indeed, from the dual flow rule~\eqref{eq:flow_dual} we have $\frac{\DD}{\DD t} H = \DD R^*(\Lambda) \sim \abs{\Lambda}^N$. For $N \to \infty$, we approach \enquote{infinitely fast} relaxation. 

On \enquote{slow} time scales we should therefore see near-infinitely fast relaxation, i.e.\ $\Sigma \in \Scal$ always, this is called \term{(local) stability}. In order to make this precise we will consider a time \emph{rescaling} and pass to the limit (this is equivalent to letting $N \to \infty$). Then, for the resulting global evolution, we obtain a rate-\emph{independent} evolution, at least where the solution process is continuous. This means that the system does not have any dynamics of its own and is purely driven by external forces.

However, it turns out that realistic concepts of solutions to rate-independent systems~\cite{MielkeTheilLevitas02,MielkeTheil04,Stefanelli09,MielkeRossiSavare09,MielkeRossiSavare12} must allow the system to \emph{jump} from one state to another instantaneously (with respect to the slow global time scale). A key question, which is also central in other recent works~\cite{Mielke02,Mielke03a,MielkeRossiSavare09,MielkeRossiSavare12,DalMasoDeSimoneSolombrino10,DalMasoDeSimoneSolombrino11}, is whether during a jump at \enquote{infinite} speed the modeling assumption of rate-independence can be upheld. Most materials in fact display rate-\emph{dependent} behavior under fast deformations and this should be reflected in the modeling.

In a way opposite to the usual approach of starting with a rate-independent flow rule and then additionally specifying what happens during fast jumps, our strategy conversely starts with a rate-\emph{dependent} flow rule and rescales time so that the slow (global) evolution becomes rate-\emph{independent} wherever it is continuous, but during fast jumps the flow can be rate-dependent, rate-independent or a mixture of both. Thus, our model has \emph{two} time parameters and hence we call our solutions \term{two-speed solutions}. Consequently, we posit the following key assumption:
\begin{quotation}
  \textit{The basic flow rule is rate-dependent, i.e.\ the fundamental dissipation potential $R$ is super-linear in the rate.}
\end{quotation}

In order to arrive at a full evolutionary model, we consider the evolution to be partitioned into segments of equal duration and describe the dynamics as we step from one time point to the next. Here, we use the two fundamental motions (purely elastic minimization and elasto-plastic relaxation) described in the previous section. In order to combine them, we make the following fundamental assumption:
\begin{quotation}
  \textit{Elastic deformations are much faster than plastic flow. Equivalently, there is infinitesimal plastic inertia.}
\end{quotation}
This is verified quite well experimentally, see for example~\cite{ArmstrongArnoldZerilli09,BenDavidEtAl14} and is also theoretically consistent (at least for metals) since the flow of dislocations is constricted to be always slower than the propagation of elastic deformation through shear waves (S-waves). This is so because the plastic drag tends to infinity as the plastic distortion rate approaches the shear wave speed, see~\cite{DeHossonyRoosMetselaar02}.

Our time-stepping scheme then progresses by alternating between the following two stages:

\begin{enumerate}[(I)]
\item \textbf{Elastic minimization:}  When the material is moved out of elastic equilibrium, it first deforms in a purely elastic fashion, where the elastic distortion jumps (instantaneously) to a minimizer of the elastic energy, completely ignoring any plastic flow effects. Consequently, there may be an \term{elastic excess}, where the stress is outside the \term{elastic stability domain}, i.e.\ the constitutively specified stress region where the system cannot plastically flow (see below). We will use the \emph{Principle of Virtual Power} (or, alternatively, a balance of forces) under the assumptions $\dot{P}=0$, $\dot{z}=0$ to derive the elastic force balance.

\item[(II)] \textbf{Elasto-plastic relaxation:} To correct the elastic excess if it exists, the material then \emph{relaxes} through an elasto-plastic flow, which is purely internal in the sense that the observable macroscopic shape of the body remains the same, i.e., $\dot{y}=0$. Again via the \emph{Principle of Virtual Power} we derive the elasto-plastic power balance and then the flow rule. The flow stops once the stress has reached the elastic stability domain.
\end{enumerate}

This splitting of the motion into the above two basic modes is the defining feature of the present model and we consider it to be physically realistic (it is also somewhat reminiscent of predictor--corrector schemes in numerics). Another benefit is that it allows for a clean modeling. In this context notice that the elastic minimality from Stage~I will potentially be destroyed by the relaxation in Stage~II, and the condition that the stress is contained in the elastic stability domain might no longer hold true after a successive Stage~I (even if the external loading does not change). Therefore, we really should iterate these stages until we have reached a \emph{fixed point}. This, however, turns out not to be necessary, since we will perform both stages for each time step in a time-stepping problem and letting the time step size go to zero will have the same effect.



\subsection{Fast flow rule} \label{sc:fastflow}

As outlined above, let $R \colon \hfrak \to [0,+\infty]$ be a given \term{rate-dependent dissipation (pseudo)potential} that we assume to be convex, lower semicontinuous, and \term{superlinear}, that is
\[
  \liminf_{\abs{\Lambda} \toup \infty}\, \frac{R(\Lambda)}{\abs{\Lambda}} = \infty.
\]
Furthermore, for $\Gamma \notin \Scal$, we want to assume that $R^*(\Gamma)$ is \emph{continuously differentiable} in $\Gamma$, expressing the fact that we model a \emph{deterministic} system, i.e.\ that the flow direction and speed are uniquely determined. Then, also $R(\Lambda)$ is continuously differentiable for all $\Lambda$ such that $\Lambda = \DD R^*(\Gamma)$ for some $\Gamma \notin \Scal$, see~\cite{Rock70CA}.

Note that additionally $R$ could depend on the current state $(\nabla y,H) \in \GL^+(d) \times \Hfrak$, but we suppress this dependency for ease of notation.

As long as $\Sigma \notin \Scal$, plastic flow is governed by our flow rule
\[
  \Sigma \in \partial R\biggl( \frac{\DD}{\DD t} H \biggr)  \quad\Longleftrightarrow\quad
  \frac{\DD}{\DD t} H \in \partial R^*(\Sigma),
\]
see~\eqref{eq:flow}, \eqref{eq:flow_dual}. Additionally, we want to express the notion that the flow stops once we have reached the elastic stability domain $\Scal$. Thus, we really use the \term{combined flow rule}
\begin{equation} \label{eq:cutoff_flow}
  \frac{\DD}{\DD t} H = \begin{cases}
    0                &\text{if $\Sigma \in \Scal$,}\\
    \DD R^*(\Sigma)  &\text{otherwise.}
  \end{cases}
\end{equation}

Clearly, we only need the dual flow potential $R^*(\Sigma)$ for $\Sigma \notin \Scal$ since the flow stops once we have reached the yield surface $\partial \Scal$. Therefore, we now change notation slightly and from here onwards denote by $R \colon \hfrak \to [0,+\infty)$ the \term{\emph{combined} dissipation potential}, for which we furthermore require the decomposition
\begin{equation} \label{eq:Rdecomp}
  R(\Gamma) = R_1(\Gamma) + R_+(\Gamma)
\end{equation}
into the following two components:
\begin{enumerate}[(i)]
  \item $R_1 \colon \hfrak \to [0,+\infty)$ is the \term{rate-independent dissipation potential}, which is equal to the support function of $\Scal$,
\[\qquad
  R_1(\Lambda) = \sigma_{\Scal}(\Lambda)
  := \sup_{\Gamma \in \Scal} \Lambda \circ \Gamma,
\]
i.e.\ the dual of the characteristic function $\chi_{\Scal}$ of $\Scal$ (which is $0$ on $\Scal$ and $+\infty$ otherwise).
  \item $R_+ \colon \hfrak \to [0,+\infty)$ is the convex, differentiable \term{residual dissipation potential}. We also assume that its dual $R_+^*$ is strictly positive, that is $R_+^*(0) = 0$ and $R_+^*(\Gamma) > 0$ for $\Gamma \neq 0$ (this is essentially a superlinearity assumption on $R_+$ and for instance satisfied if $\liminf_{t\to 0}\, R_+(t\Lambda)/t = 0$).
\end{enumerate}

It turns out that the decomposition~\eqref{eq:Rdecomp} implies
\begin{equation} \label{eq:Rconj_structure}
  R^*(\Sigma)
  = \inf_{\Gamma \in \Scal} R_+^*(\Sigma-\Gamma).
\end{equation}
Indeed, we can compute, using the inf-convolution $(f \infc g)(x) := \inf_z [f(x-z)+g(z)]$ and the associated duality rule $[f \infc g]^* = f^* + g^*$ (see Theorem~16.4 of~\cite{Rock70CA}), that
\[
  R^*(\Sigma) = [R_1 + R_+]^*(\Sigma)
  = [R_1^* \infc R_+^*](\Sigma)
  = \inf_{\Gamma \in \Scal} R_+^*(\Sigma - \Gamma).
\]
In particular, using the strict positivity of $R_+^*$, we have for the \term{dual \emph{combined} dissipation potential} $R^* \colon \hfrak^* \to [0,+\infty]$ the property that
\[
  R^*(\Sigma) = 0  \quad\text{if and only if}\quad
  \Sigma \in \Scal.
\]
Consequently, $R^*$ by itself already determines $\Scal$ and the flow rule for our new $R$ is the combined flow rule~\eqref{eq:cutoff_flow}.

On the other hand, assume that we are given $\Scal \subset \hfrak^*$ with $0 \in \Scal$ and a convex, differentiable \term{dual residual dissipation potential} $R_+^* \colon \hfrak^* \to [0,+\infty)$ with the strict positivity property from above. Then, we can define $R^*$ (and hence by duality $R$) through~\eqref{eq:Rconj_structure} and we obtain for $R$ that
\[
  R = [R^*]^*
    = \bigl[\chi_{\Scal} \infc R_+^*\bigr]^*
    = R_1 + R_+
\]
where we have set $R_1 := \sigma_{\Scal} = \chi_{\Scal}^*$ and $R_+ := R_+^{**} := [R_+^*]^*$. Thus, the decomposition above holds, where we also note that $R_1 \geq 0$ because of $0 \in \Scal$ and $R_+ = R_+^{**} \geq 0$ because of $R_+^*(0) = 0$.

\subsection{Associative flow potentials}

An important special case, which we discuss for further illustration, is a material that is \term{associative}. While there is considerable disagreement in the literature over what exactly constitutes an associative flow rule, we here understand it to mean that there exists a proper, lower semicontinuous, convex, and strictly increasing $\zeta_+^* \colon [0,+\infty) \to [0,+\infty)$ and a norm $\norm{\frarg}_+^*$ on $\hfrak^*$ such that
\[
  R_+^*(\Gamma) = \zeta_+^*(\norm{\Gamma}_+^*),  \qquad
  \Gamma \in \hfrak^*.
\]
The utility of these assumptions is that they allow to simplify the definition (cf.~\eqref{eq:Rconj_structure}) of the dual combined dissipation potential $R^* \colon \hfrak^* \to [0,+\infty)$ to
\[
  R^*(\Gamma)
  = \begin{cases}
    0   & \text{if $\Gamma \in \Scal$,} \\
    \zeta_+^*(\dist_{\norm{\frarg}_+^*}(\Gamma,\Scal)
        & \text{if $\Gamma \notin \Scal.$}
  \end{cases}
\]
where $\dist_{\norm{\frarg}_+^*}(\Gamma,\Scal) := \inf_{\Gamma' \in \Scal} \norm{\Gamma-\Gamma'}_+^*$ is the \term{distance function} with respect to the norm $\norm{\frarg}_+^*$. In this case, all flow rates are normal to the elastic stability domain, that is,
\[
  \partial R^*(\Gamma) \subset \Nrm_{\Scal}(\Gamma)  \qquad
  \text{for all $\Gamma \in \partial \Scal$.}
\]
where $\Nrm_{\Scal}(\Gamma)$ denotes the normal cone to $\Scal$ at $\Gamma \in \partial \Scal$, i.e.\
\[
  \Nrm_{\Scal}(\Gamma) := \setb{ \Lambda \in \hfrak }{ \text{$\Lambda \circ (\Gamma'-\Gamma) \leq 0$ for all $\Gamma' \in \Scal$} }.
\]
This is related to the \emph{Maximum Plastic Work Principle}, which is a strengthening of the Second Law of Thermodynamics, see p.~57--59 in~\cite{HanReddy13}.

\begin{example}
A common choice for a rate-dependent dissipation potential $R \colon \mathfrak{sl}(d) \to \R$ (recall that $\mathfrak{sl}(d)$ contains all trace-free $(d \times d)$-matrices), is a Mises power law, see for instance Section~101 in~\cite{GurtinFriedAnand10}, that is, the restriction of 
\[
  \hat{R}(V) := \frac{\kappa}{1+1/N} \abs{V}^{1+1/N}
\]
to the complement of the elastic stability domain
\[
  \Scal = \setb{ V \in \mathfrak{sl}(d)^* }{ \hat{R}^*(V) \leq \kappa/(1+N) } = B_{\mathfrak{sl}(d)^*}(0,\kappa) \subset \mathfrak{sl}(d)^*.
\]
Here, $\kappa > 0$ and $N > 1$. Since for $V \in \mathfrak{sl}(d)^*$,
\[
  \hat{R}^*(V) = \frac{\kappa}{1+N} \biggl(\frac{\abs{V}}{\kappa}\biggr)^{1+N},
\]
we may compute
\[
  R_1(V) = \sigma_\Scal(V) = \kappa \abs{V}  \qquad\text{and}\qquad
  R_+(V) = \kappa \biggl[ \frac{\abs{V}^{1+1/N}}{1+1/N} + \frac{1}{1+N} - \abs{V} \biggr]^+,
\]
where $[s]^+ = \max(s,0)$. Then, indeed $R|_{\mathfrak{sl}(d) \setminus \Scal} = \hat{R}|_{\mathfrak{sl}(d) \setminus \Scal}$.
\end{example}

\subsection{Relaxation paths} \label{sc:paths}

Let $F_0 \in \GL^+(d)$, $(P_0,z_0) \in \Hfrak$ such that $\Sigma(F_0,P_0,z_0) \notin \Scal$. Then the system is driven towards relaxation by following a \term{relaxation path} $\gamma \colon [0,\tau_\gamma] \to \Hfrak$, where $\tau_\gamma>0$ may depend on the path $\gamma$, until relaxation is achieved when $\Sigma(F_0,\gamma(\tau_\gamma)) \in \partial\Scal$. We define $\gamma$ through
\[
  \left\{\begin{aligned}
    \gamma(0)                   &= (P_0,z_0),  \\
    \frac{\DD}{\DD t} \gamma(t) &\in \partial R^* \bigl(\Sigma(F_0,\gamma(t)) \bigr), \quad t \in [0,\tau_\gamma), \\
    \Sigma(F_0,\gamma(t))       &\notin \Scal \quad\text{for}\quad t \in [0,\tau_\gamma), \\
    \Sigma(F_0,\gamma(\tau_\gamma))  &\in \partial \Scal.
  \end{aligned}\right.
\]
Note that $\frac{\DD}{\DD t} \gamma = (\frac{\DD}{\DD t} P,\dot{z})$, where $\gamma(t) = (P(t), z(t))$.

Assuming that this path exists and is unique, we denote its endpoint by $Y(f_0,P_0,z_0)$. If we also set $Y(F_0,P_0,z_0) := (P_0,z_0)$ if already $\Sigma(F_0,P_0,z_0) \in \Scal$, then $Y$ can be considered a function $Y \colon \GL^+(d) \times \Hfrak \to \Hfrak$ with the property that
\[
  \Sigma(F_0,Y(P_0,z_0)) \in \Scal  \qquad
  \text{for all $(F_0,P_0,z_0) \in \GL^+(d) \times \Hfrak$.} 
\]

\subsection{Time rescaling and time-stepping evolution}

Conceptually, rate-independence is of course not a physical property of a system, but a \emph{mathematical} rescaling limit. So, consider an elasto-plastic process $u(t) = (y(t),P(t),z(t))$ with external loading $f = f(t)$, where $t \in [0,T)$. Then, the basic assumption in this work is that the system evolves according to a (fast) flow rule as defined above. However, as discussed before, if $f$ were to be held constant at some point in time, the system would settle very quickly into a rest state until the external loading changes and the system is pushed out of equilibrium. The traditional rate-independent modeling is built upon the assumption that only this global movement is interesting and the fast \enquote{relaxation} movements towards a rest state can be neglected, at least if the system does not jump to a far-away state in an instant.

This idealized situation is mathematically expressed through \emph{rescaling}: Define for small $\lambda > 0$ the process $\tilde{u}_\lambda(\theta)$ as a solution of the fast dynamics for the slower external loading $\tilde{f}_\lambda(\theta) = f(\lambda \theta)$, where now $\theta \in [0,T/\lambda]$. It is, however, inconvenient to deal with changing time intervals and so we reparametrize the whole process by setting $u_\lambda(t) := \tilde{u}_\lambda(t/\lambda)$, $t \in [0,T)$. This has the effect of keeping the external loading at constant speed, but accelerating the response dynamics of the system by the factor $1/\lambda$.

Letting $\lambda \todown 0$, we call the limit process $u_0(t) := \lim_{\lambda\todown 0} u_\lambda(t)$ the \term{rate-independent limit}. It models the situation where the relaxation behavior of the system is infinitely fast and the system remains completely stationary if the external loading does not change. However, this acceleration is only due to a rescaling of time, which does not affect the dissipative energetics. So, effectively, this new system has a different (rescaled) dissipation law to reflect the rescaling in time. More concretely, the \enquote{fast} relaxation paths $\gamma_\lambda$ (corresponding to the $\gamma$'s from the previous section) for $u_\lambda$ have to modified to read
\[
  \lambda \frac{\DD}{\DD t} \gamma_\lambda(t)
  \in \partial R^* \bigl( \Sigma(F_0,\gamma_\lambda(t)) \bigr), \quad t \in [0,\lambda\tau_\gamma).
\]
In the following we will develop an approximative, time-stepping, scheme whose solution reflects this sped-up response.

For reasons of consistency, we assume that the initial values $u_\start = (y_\start,H_\start)$ where the diffeomorphism $y_\start \colon \Omega \to \R^d$ and $H_\start \colon \Omega \to \Hfrak$ constitute a rest state, i.e.\
\[
  y_\start \quad \text{minimizes} \quad
  \hat{y} \mapsto \Ecal[0,\hat{y},H_\start]
  \qquad\text{and}\qquad
  \Sigma(u_\start) \in \Scal.
\]
Here, recall the definition of $\Ecal$ from~\eqref{eq:Ecal}. This condition is imposed to ensure that the system does not immediately jump and we should have chosen the endpoint of the jump as initial value.

Let $N \in \N$ and divide the time interval $[0,T)$ into an equidistant partition 
\[
  0 = t_0^N < t_1^N < \cdots < t_{2^N}^N = T,  \qquad
  t_k = k 2^{-N}T  \quad\text{($k = 0, \ldots, 2^N$).}
\]
We also need the \term{rescaling coefficient} $\lambda^N > 0$, which describes the time-rescaling. We require that that $\lambda^N \todown 0$ as $N \to \infty$ \enquote{more slowly} than the partition, that is,
\[
  \lambda^N 2^N \toup \infty  \qquad \text{as $N \to \infty$.}
\]
This constraint expresses that the elastic minimization is relatively faster than the elasto-plastic relaxation.

The \term{time-stepping evolution} $u^N = (y^N,P^N,z^N)$ consists of the component functions $y^N \colon [0,T) \times \Omega \to \R^d$, $P^N \colon [0,T) \times \Omega \to \Pfrak$, $z^N \colon [0,T) \times \Omega \to \Zfrak$. In every consecutive time period $[t_k,t_{k+1})$ the discrete system in turn executes the following two \emph{stages} (illustrated in Figure~\ref{fig:twostage}):

\begin{figure}[tb]
\begin{center}
\includegraphics[scale=0.9]{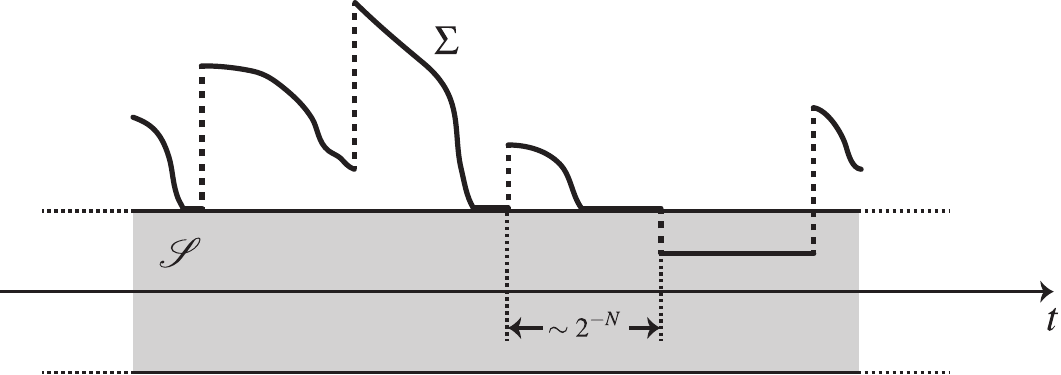}
\caption{The two-stage time-stepping scheme.} 
\label{fig:twostage}
\end{center}
\end{figure}



\begin{figure}[tb]
\begin{center}
\includegraphics[scale=0.9]{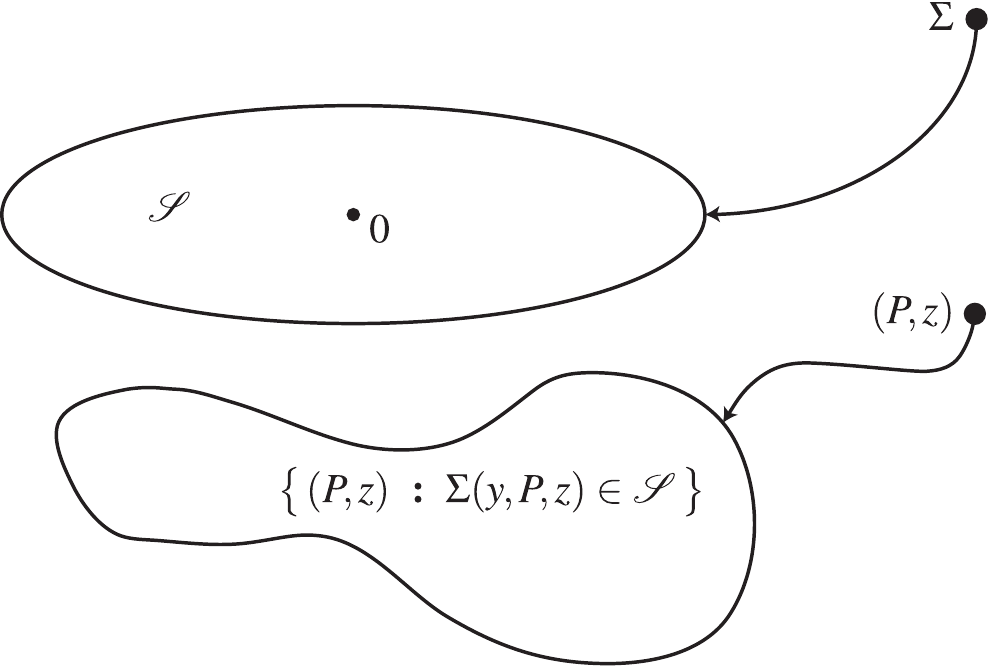}
\caption{The elasto-plastic relaxation flow.} 
\label{fig:relaxflow}
\end{center}
\end{figure}

\begin{enumerate}[(I)]
\item \textbf{Elastic minimization:} First, we let the system deform elastically \emph{as if there was no plastic flow rule present}. Let the diffeomorphism $y_* \colon \Omega \to \R^d$ be such that (holding $H^N(t_k) := (P^N(t_k),z^N(t_k))$ fixed)
\[\qquad
  y_* \quad \text{minimizes} \quad
  \hat{y} \mapsto \Ecal[t_k,\hat{y},H^N(t_k)]
\]
over all diffeomorphisms $\hat{y} \colon \Omega \to \R^d$ that also satisfy the prescribed boundary conditions. We assume that the system's total deformation jumps immediately to $y_*$:
\[\qquad
  y^N(t) := y_*,  \qquad t \in [t_k,t_{k+1}).
\]

\item \textbf{Elasto-plastic relaxation:} Whenever necessary, the system will now exchange elastic for plastic distortion to restore the requirement that the generalized stress lies within the elastic stability domain. We define the internal state at a fixed $x \in \Omega$ as
\[\qquad
  H^N(t,x) := (P^N(t,x),z^N(t,x)) \in \Hfrak,  \qquad t \in [t_k,t_{k+1}).
\]
Then, for the evolution of $H^N_x$ we consider two cases, depending on the value of $F_0 := \nabla y_*$:

\begin{enumerate}[(a)]
  \item If $\Sigma(F_0,H^N(t_k,x)) \in \Scal$, then set (no flow)
\[\qquad\qquad
  H^N(t,x) := H^N(t_k,x),  \qquad t \in [t_k,t_{k+1}).
\]
  \item Otherwise, the system needs to relax through elasto-plastic flow, for which it employs a relaxation path $\gamma$ as in Section~\ref{sc:paths} with $F_0 = \nabla y_*(x) = \nabla y^N(t_k,x)$ (for our fixed $x$). Let $\gamma \colon [0,\tau_\gamma] \to \Hfrak$ be the relaxation path starting at $H^N(t_k,x)$, which we consider to be constantly extended to $\gamma \colon [0,+\infty) \to \Hfrak$ and set
\[\qquad\qquad
  H^N(t,x) := \gamma \biggl( \frac{t-t_k}{\lambda^N} \biggr),  \qquad
  t \in [t_k,t_{k+1}).
\]
Note that it is possible that, depending on the value of $\lambda^N$, the relaxation path may be shorter than the full interval (if $\tau_\gamma \lambda^N \leq 2^{-N}T$
) or longer, in which case the relaxation is not complete by the next time point $t_{k+1}$. Furthermore, since we assumed that all relaxations are internal, during the relaxation the full deformation does not change. See Figure~\ref{fig:relaxflow} for an illustration of the relaxation.
\end{enumerate}
Thus, we have the \term{effective flow rule}
\[\qquad\qquad
  \Sigma(u^N(t,x)) \in \partial R \biggl( \lambda^N \frac{\DD}{\DD t} H^N(t,x) \biggr),
  \qquad (t,x) \in (t^N_k,t^N_{k+1}) \times \Omega.
\]
Note that only if $\tau_\gamma \lambda^N \leq 2^{-N}T$, at the end of the relaxation path it holds that
\[\qquad
  \Sigma(u^N(t_{k+1},x)) \in \Scal,
\]
hence the system has reached an elastically allowable state. Of course, in this process, $y^N(t_k)$ does not necessarily retain the minimization property from stage~(I).
\end{enumerate}

We iterate this scheme for $k = 1, \ldots, 2^N$ and call the resulting $u^N = (y^N,H^N)$ the \term{time-stepping evolution} at level $N$.

\section{Limit passage and two-speed solutions} \label{sc:limit}

Assume we have a sequence of time-stepping solutions $u^N \colon [0,T) \times \Omega \to \R^d \times \Hfrak$. In this section we will consider heuristics of the limit passage $N \to \infty$ in order to understand the continuous-time behavior. This can be seen as a \enquote{blueprint} of a future fully rigorous mathematical analysis. We do not state all the precise assumptions, but indicate key requirements as we go along.

As the basis of all of the following we require:
\begin{enumerate}
  \item[(A1)]\quad $\displaystyle u^N \to u_0 = (y_0,P_0,z_0) \colon [0,T) \times \Omega \to \R^d \times \Hfrak$ in a sufficiently good sense.
\end{enumerate}
This would have to be made precise in a rigorous treatment, we here simply assume that the convergence is \enquote{good enough} to make all the following arguments work.

We will distinguish two types of points $t_0 \in [0,T)$: regular and singular points. At \term{regular points}, we assume the bound
\begin{equation} \label{eq:regpoint_ass}
  \int_\Omega \absBB{\frac{\DD}{\DD t} H^N(x,t)} \dd x  \quad\text{is uniformly (in $N,t$) bounded for $t \in [t_0-\eps,t_0+\eps]$.} 
\end{equation}
Here, $\eps > 0$ may depend on $t_0$, but we suppress this in the notation. All other points are called \term{singular points}. Accordingly, at regular points the internal variables do not change too fast, whereas at singular points $u^N$ may develop a jump.

We start by making two preliminary observations: First, from the (fast) flow inclusion in the elasto-plastic relation stage,
\[
  \Sigma(u^N(t,x)) \in \partial R \biggl( \lambda^N \frac{\DD}{\DD t} H^N(t,x) \biggr),
  \qquad (t,x) \in (t^N_k,t^N_{k+1}) \times \Omega,
\]
we estimate for $\tau \in (t^N_k,t^N_{k+1})$ that
\begin{align}
  - \frac{\di}{\di t} \Wcal[u^N(t)]
  &= \frac{1}{\lambda^N} \int_\Omega \Sigma(u^N(t)) \circ \lambda^N \frac{\DD}{\DD t} H^N(t) \dd x  \notag\\
  &= \frac{1}{\lambda^N} \int_\Omega R \biggl( \lambda^N \frac{\DD}{\DD t} H^N(t) \biggr) + R^* \bigl( \Sigma(u^N(t)) \bigr) \dd x  \notag\\
  &\geq \int_\Omega R_1 \biggl( \frac{\DD}{\DD t} H^N(t) \biggr) \dd x  
  \geq 0. \label{eq:flow_apriori}
\end{align}
Here we used that if $\Gamma \in \partial R(\Lambda)$, then $\Lambda \circ \Gamma = R(\Lambda) + R^*(\Gamma)$, this is Fenchel's Theorem, see for instance Theorem~23.5 in~\cite{Rock70CA} and $R_+ \geq 0$, $R^* \geq 0$. Note that here and in the following we left out the $x$-argument of the functions under the integral.

Second, for $H \colon [0,T) \times \Omega \to \Hfrak$, we define the \term{(rate-independent) $R_1$-dissipation} on the subinterval $[s,t] \subset [0,T)$ as follows:
\[
  \Diss_1(H;[s,t]) := \sup\, \setBBB{ \sum_{k=1}^N \Dcal_0(H(\tau_{k-1}), H(\tau_k)) }{ s = \tau_0 < \tau_1 < \cdots < \tau_N = t, \; N \in \N },
\]
where for $G_0,G_1 \in \Hcal$,
\begin{align*}
  \Dcal_0(G_0,G_1) &:= \int_\Omega d_{R_1}(G_0(x),G_1(x)) \dd x, \\
  d_{R_1}(H_0,H_1) &:= \inf\, \set{ \int_0^1 R_1 \biggl( \frac{\DD}{\DD \tau} \gamma(\tau) \biggr) \dd \tau }{ \begin{aligned}&\text{$\gamma \colon [0,1] \to \Hfrak$ differentiable,}\\&\text{$\gamma(0) = H_0$, $\gamma(1) = H_1$}\end{aligned} }.
\end{align*}

If $H_0$ is continuously differentiable, then we have for the $R_1$-dissipation on the subinterval $[s,t] \subset [0,T)$ that
\[
  \Diss_1(H_0;[s,t]) = \int_s^t \int_\Omega R_1 \biggl( \frac{\DD}{\DD \tau} H_0(\tau) \biggr) \dd x \dd \tau.
\]
We assume here that under a sufficiently strong convergence, $\Diss_1$ is \term{lower semicontinuous}, that is, if $H_j \to H$ (with respect to our sufficiently strong convergence), then
\[
  \Diss_1(H;[s,t]) \leq \liminf_{j\to\infty}\, \Diss_1(H_j;[s,t]).
\]
This can be justified (even for not so strong convergences $H_j \to H$) by appealing to the \emph{convexity} of $R_1$.



Also, from~\eqref{eq:flow_apriori}, we get
\[
  \int_0^T \int_\Omega R_1 \biggl( \frac{\DD}{\DD \tau} H^N(\tau) \biggr) \dd x \dd \tau \leq \Wcal[u^N(0)] - \Wcal[u^N(T)] \leq \Wcal[u_\start] < \infty.
\]
Therefore,
\begin{equation} \label{eq:Diss_uN_bounded}
  \Diss_1(H^N;[0,T)) \leq C  \quad\text{with $C > 0$ uniform in $N$.}
\end{equation}
In particular, the $H^N$ are of uniformly bounded variation.

\subsection{Regular points}

Let $t_0 \in [0,T)$ be a regular point. Assume
\begin{enumerate}
  \item[(A2)]\quad $\displaystyle \Ecal[t_0,\frarg]$ is lower semicontinuous with respect to the convergence $u^N \to u_0$.
\end{enumerate}
Then (the first inequality being our notion of lower semicontinuity),
\begin{align*}
  \Ecal[t_0,u_0(t_0)] &\leq \liminf_{N\to\infty} \Ecal[t_0,u^N(t_0)] \\
  &= \liminf_{N\to\infty} \biggl( \Ecal[t^N_*,u^N(t^N_*)] + \int_{t^N_*}^{t_0} \frac{\di}{\di \tau} \Ecal[\tau,u^N(\tau)] \dd \tau \biggr),
\end{align*}
where $t^N_*$ is the $t^N_k$ such that $t_0 \in [t^N_k, t^N_{k+1})$. The integral can be estimated as (notice that $y^N$ is constant in the interval $ [t^N_*,t_0)$ by construction)
\begin{align*}
  \absBB{\int_{t^N_*}^{t_0} \frac{\di}{\di \tau} \Ecal[\tau,u^N(\tau)] \dd \tau}
  &\leq \absBB{\int_{t^N_*}^{t_0} \int_\Omega \Sigma(u^N(\tau)) \circ \frac{\DD}{\DD \tau} H^N(\tau) \dd x \dd \tau} \\
  &\qquad + \absBB{\int_{t^N_*}^{t_0} \int_\Omega \dot{f}(\tau) \cdot y^N(\tau) \dd x \dd \tau} \\
  &\leq C 2^{-N} \to 0  \qquad\text{as $N \to \infty$.}
\end{align*}
Here we used $\DD_H W(u^N) \cdot \frac{\di}{\di t} H^N = -\Sigma(u^N) \circ \frac{\DD}{\DD t} H^N$ together with the assumption~\eqref{eq:regpoint_ass} as well as
\begin{enumerate}
  \item[(A3)]\quad $\displaystyle \absb{\Sigma(u^N(t,x))} \leq C$ for all $(t,x) \in [0,T] \times \Omega$ uniformly in $N, t$,
  \item[(A4)]\quad $\displaystyle \int_\Omega \abs{y^N(t)} \dd x \leq C$ uniformly in $N, t$,
  \item[(A5)]\quad $\displaystyle \abs{\dot{f}(t,x)} \leq C$ uniformly in $N, t, x$.
\end{enumerate}
Now, at $t^N_*$, our $y^N(t^N_*)$ is a minimizer of $\hat{y} \mapsto \Ecal[t^N_*,\hat{y},H^N(t^N_*)]$ and so, for any $\hat{y} \colon \Omega \to \R^d$,
\[
  \Ecal[t_0,u_0(t_0)] \leq \liminf_{N\to\infty} \Ecal[t^N_*,u^N(t_*)]
  \leq \liminf_{N\to\infty} \Ecal[t^N_*,\hat{y},H^N(t^N_*)]
  = \Ecal[t_0,\hat{y},H_0(t_0)],
\]
where for the last limit passage we need to assume that the convergence of $H^N \to H_0$ is strong enough and that $H^N(t^N_*) \to H_0(t_0)$ such that $\Ecal[t^N_*,\hat{y},H^N(t^N_*)] \to \Ecal[t_0,\hat{y},H_0(t_0)]$. Thus,
\begin{enumerate}
  \item[(E1$_{\mathrm{reg}}$)]\quad $\displaystyle
  y_0(t_0) \quad \text{minimizes} \quad
  \hat{y} \mapsto \Ecal[t_0,\hat{y},H_0(t_0)]  \qquad
  \text{for regular $t_0 \in [0,T)$.}$
\end{enumerate}

Next, we turn to the stability. By the definition of time-stepping solutions, we have from Section~\ref{sc:paths} that
\[
  \Sigma(u^N(t_0)) \in \partial R \biggl( \lambda^N \frac{\DD}{\DD t} H^N(t_0) \biggr)
  = \partial R_1 \biggl( \lambda^N \frac{\DD}{\DD t} H^N(t_0) \biggr)
  + \DD R_+ \biggl( \lambda^N \frac{\DD}{\DD t} H^N(t_0) \biggr).
\]
Hence, again using the identity $\Lambda \circ \Gamma = R(\Lambda) + R^*(\Gamma)$ if $\Gamma \in \partial R(\Lambda)$ (see Theorem~23.5 in~\cite{Rock70CA}), we get 
\begin{align}
  \Diss(u^N;[t_0-\eps,t_0+\eps])
  &= \int_{t_0-\eps}^{t_0+\eps} \int_\Omega \Sigma(u^N(\tau)) \circ \frac{\DD}{\DD \tau} H^N(\tau) \dd x \dd \tau  \notag\\
  &= \frac{1}{\lambda^N} \int_{t_0-\eps}^{t_0+\eps} \int_\Omega R \biggl( \lambda^N \frac{\DD}{\DD \tau} H^N(\tau) \biggr) + R^*\bigl( \Sigma(u^N(\tau)) \bigr) \dd x \dd \tau  \notag\\
  &\geq \frac{1}{\lambda^N} \int_{t_0-\eps}^{t_0+\eps} \int_\Omega R^*\bigl( \Sigma(u^N(\tau)) \bigr) \dd x \dd \tau  \label{eq:Diss_stab_est}
\end{align}
since $R \geq 0$. On the other hand, we assumed that around the regular point $\Sigma(u^N(\tau))$ is uniformly (in $N$ and $x$) bounded. Then, as $\abs{\Lambda} \leq C R_1(\Lambda)$,
\begin{align*}
  \Diss(u^N;[t_0-\eps,t_0+\eps])
  &\leq C \int_{t_0-\eps}^{t_0+\eps} \int_\Omega \absBB{\frac{\DD}{\DD \tau} H^N(\tau)} \dd x \dd \tau \\
  &\leq C \int_{t_0-\eps}^{t_0+\eps} \int_\Omega R_1 \biggl(\frac{\DD}{\DD \tau} H^N(\tau) \biggr) \dd x \dd \tau
  \leq C
\end{align*}
by~\eqref{eq:Diss_uN_bounded}, with $C > 0$ independent of $N$. Combining, we get that
\[
  \int_{t_0-\eps}^{t_0+\eps} \int_\Omega R^*\bigl( \Sigma(u^N(\tau)) \bigr) \dd x \dd \tau
  \leq C\lambda^N \to 0  \quad\text{as $N \to \infty$.}
\]
In the limit we have ($R^* \geq 0$)
\[
  R^*\bigl( \Sigma(u_0(\tau,x)) \bigr) = 0  \quad\text{for $\tau \in [t_0-\eps,t_0+\eps]$, $x \in \Omega$.}
\]
Since $R^*(\Gamma)$ is zero if and only if $\Gamma \in \Scal$ by~\eqref{eq:Rconj_structure}, this implies the stability condition
\begin{enumerate}
  \item[(E2)]\quad $\displaystyle
  \Sigma(u_0(t_0,x)) \in \Scal \quad\text{for $t_0$ regular, $x \in \Omega$.}$
\end{enumerate}

\subsection{Singular points}

At singular points $t_0 \in [0,T)$ instead of~\eqref{eq:regpoint_ass} we only require
\begin{enumerate}
  \item[(A6)]\quad $\displaystyle \lambda^N \int_\Omega \absBB{\frac{\DD}{\DD t} H^N(x,t)} \dd x \quad\text{is uniformly (in $N,t$) bounded for $t \in [0,T)$.}$
\end{enumerate}
This can be justified as follows: If our minimizers in Stage~(I) always have enough regularity (i.e.\ enough derivatives), then the speed of the unrescaled elasto-plastic relaxation flow remains bounded. Since we speed up the flow by a factor of $1/\lambda^N$, the assumption~(A6) is then realistic. Also, the stress $\Sigma$ is uniformly bounded everywhere, i.e.~(A3) holds. In a forthcoming work on geometrically linear rate-independent systems~\cite{Rindler15TwoSpeedRI}, (A3),~(A6) can be proved rigorously (albeit for the quadratic $\Lrm^2$-norm).

In the singular case, we need to \emph{rescale} our processes $u^N$ around $t_0$ as follows: Set
\begin{equation} \label{eq:uN_rescale_lim}
  u^N(t_0,\theta) := u^N(t_0 + \lambda^N \theta),  \qquad
  \theta \in (-\infty,+\infty).
\end{equation}
Since then
\[
  \int_\Omega \absBB{\frac{\DD}{\DD \theta} H^N(x,t_0 + \lambda^N \theta)} \dd x
\]
is uniformly in $N, \theta$ bounded by the chain rule for the referential derivative, we may assume that
\[
  u^N(t_0,\theta) \to u(t_0,\theta) \;\;\text{in a sufficiently strong way}
  \quad\text{and}\quad
  u(t_0,\pm \infty) = u_0(t_0\pm).
\]
For the minimization property, we can argue in a similar way as we did at regular points: It holds that
\[
  \Ecal[t_0,u(t_0,\theta)] \leq \liminf_{N\to\infty} \biggl( \Ecal[t^N_*,u^N(t^N_*)] + \int_{t^N_*}^{t_0+\lambda^N \theta} \frac{\di}{\di \tau} \Ecal[\tau,u^N(\tau)] \dd \tau \biggr),
\]
where $t^N_*$ is the $t^N_k$ such that $t_0 + \lambda^N \theta \in [t^N_k, t^N_{k+1})$. We estimate
\begin{align*}
  \absBB{\int_{t^N_*}^{t_0+\lambda^N \theta} \frac{\di}{\di \tau} \Ecal[\tau,u^N(\tau)] \dd \tau}
  &\leq \absBB{\int_{t^N_*}^{t_0+\lambda^N \theta} \int_\Omega \Sigma(u^N(\tau)) \circ \frac{\DD}{\DD \tau} H^N(\tau) \dd x \dd \tau} \\
  &\qquad + \absBB{\int_{t^N_*}^{t_0+\lambda^N \theta} \int_\Omega \dot{f}(\tau) \cdot y(\tau) \dd x \dd \tau} \\
  &\leq C \biggl( \frac{2^{-N}}{\lambda^N} + 2^{-N} \biggr) \\
  &\to 0  \qquad\text{as $N \to \infty$}
\end{align*}
by our choice of the $\lambda^N$ as going to zero more slowly than $2^{-N}$ and also using~(A4),~(A5).

As before we get for any $\hat{y} \colon \Omega \to \R^d$ that
\[
  \Ecal[t_0,u(t_0,\theta)] \leq \liminf_{N\to\infty} \Ecal[t^N_*,u^N(t_*)]
  \leq \liminf_{N\to\infty} \Ecal[t^N_*,\hat{y},H^N(t^N_*)]
  = \Ecal[t_0,\hat{y},H(t_0,\theta)].
\]
In the last limit we use that
\begin{align*}
  \int_\Omega \absb{H^N(t_0 + \lambda^N \theta) - H^N(t^N_*)} \dd x
  &\leq \int_{t^N_*}^{t_0+\lambda^N \theta} \int_\Omega \absBB{\frac{\DD}{\DD \tau} H^N(\tau)} \dd x \dd \tau \\
  &\leq C \frac{2^{-N}}{\lambda^N} \to 0  \qquad\text{as $N \to \infty$}
\end{align*}
and further assumed continuity properties of $\Ecal$. Thus,
\begin{enumerate}
  \item[(E1$_{\mathrm{sing}}$)]\quad $\displaystyle
  y(t_0,\theta) \quad \text{minimizes} \quad
  \hat{y} \mapsto \Ecal[t_0,\hat{y},H(t_0,\theta)]  \qquad
  \text{for singular $t_0 \in [0,T)$.}$
\end{enumerate}

We briefly indicate why no stability can be expected on the transients $u(t_0,\theta)$: Here, because of the necessary rescaling, instead of~\eqref{eq:Diss_stab_est} we can only expect
\begin{align*}
  \Diss(u^N;[t_0-\lambda^N\eps,t_0+\lambda^N\eps])
  \geq \int_{-\eps}^{\eps} \int_\Omega R^*\bigl( \Sigma(u^N(t_0+\lambda^N\theta)) \bigr) \dd x \dd \theta.
\end{align*}
Now, the left hand side does not necessarily vanish (if $t_0$ is a jump point) and we have \enquote{lost} the pre-factor $\lambda^N$. Thus, we cannot conclude that $R^*(\Sigma(u(t_0,\theta))) = 0$. Consequently, the non-stability $\Sigma(u(t_0,\theta)) \notin \Scal$ is possible. This is also to be expected: On fast transients, the stability does not necessarily hold. For instance, a material that has elastically \enquote{snapped} into a new state, will need to elasto-plastically relax and during this relaxation process the stability is not (yet) valid.

%

\subsection{Energy balance}

We now turn to the energy balance. First we consider the discrete situation: Let $t \in [t^N_k,t^N_{k+1})$, where $k \in \{0,1,\ldots,2^N-1\}$. From the flow inclusion in the elasto-plastic relaxation phase (recall that $\frac{\DD}{\DD \tau} H = 0$ once we have reached the stability domain), we get
\begin{align}
  \Wcal[u^N(t)] - \Wcal[u^N(t^N_k)]
  &= \int_{t^N_k}^t \frac{\di}{\di \tau} \Wcal[u^N(\tau)] \dd \tau  \notag\\
  &= - \int_{t^N_k}^t \Sigma(u^N(\tau)) \circ \frac{\DD}{\DD \tau} H^N(\tau) \dd \tau  \notag\\
  &= -\Diss(u^N;[t^N_k,t])  \label{eq:step_est_last}
\end{align}
since $y^N(\tau)$ is constant on $[t^N_k,t^N_{k+1})$. This also extends to $t = t^N_{k+1}-$ (left limit). For any $l = 0,\ldots,k-1$ the elastic minimization in Stage~(I) and $y^N(t^N_{l+1}-) = y^N(t^N_l)$, $H^N(t^N_{l+1}-) = H^N(t^N_{l+1})$, gives
\begin{align*}
  \Ecal[t^N_{l+1},u^N(t^N_{l+1})] &\leq \Ecal[t^N_{l+1},u^N(t^N_{l+1}-)] \\
  &= \Wcal[u^N(t^N_l)] - \Diss(u^N;[t^N_l,t^N_{l+1})) - \int_\Omega f(t^N_{l+1}) \cdot y^N(t^N_l) \dd x \\
  &= \Ecal[t^N_l,u^N(t^N_l)] - \Diss(u^N;[t^N_l,t^N_{l+1})) - \int_{t^N_l}^{t^N_{l+1}} \int_\Omega \dot{f}(\tau) \cdot y^N(\tau) \dd x \dd \tau.
\end{align*}
Summing up this telescopic identity and using the additivity of $\Diss$,
\[
  \Ecal[t^N_k,u^N(t^N_k)]
  \leq \Ecal[0,u_\start] - \Diss(u^N;[0,t^N_k)) - \int_0^{t^N_k} \int_\Omega \dot{f}(\tau) \cdot y^N(\tau) \dd x \dd \tau,
\]
where $t_k^N$ is chosen such that $t \in [t_k^N,t_{k+1}^N)$. Then, using~\eqref{eq:step_est_last} again we get the \term{time-stepping upper energy bound}
\begin{equation} \label{eq:step_energy_upper}
  \Ecal[t,u^N(t)] \leq \Ecal[0,u_\start] - \Diss(u^N;[0,t)) - \int_0^t \int_\Omega \dot{f}(\tau) \cdot y^N(\tau) \dd x \dd \tau.
\end{equation}

We next pass to the (lower) limit in $\Diss(u^N;[0,t))$, which we will only perform for $t = T$ for ease of notation. Assume that $S$ is an open subset of $[0,T)$ that contains a $\eta$-neighborhood of all singular points $t_0$, i.e.\ those points where~\eqref{eq:uN_rescale_lim} holds. Then,
\[
  \Diss(u^N;[0,T)) = \Diss(u^N;[0,T)\setminus S) + \Diss(u^N;S).
\]
For the first, continuous, part we have
\begin{align*}
  \Diss^N(u^N;[0,T)\setminus S) &= \int_{[0,T)\setminus S} \int_\Omega \Sigma(u^N(\tau)) \circ \frac{\DD}{\DD \tau} H^N(\tau) \dd x \dd \tau  \\
  &= \frac{1}{\lambda^N} \int_{[0,T)\setminus S} \int_\Omega R \biggl( \lambda^N \frac{\DD}{\DD \tau} H^N(\tau) \biggr) + R^* \bigl( \Sigma(u^N(\tau)) \bigr) \dd x \dd \tau  \\
  &\geq \int_{[0,T)\setminus S} \int_\Omega R_1 \biggl( \frac{\DD}{\DD \tau} H^N(\tau) \biggr) \dd x \dd \tau  \\
  &=: \Diss_1(u^N;[0,T)\setminus S).
\end{align*}
For the singular part, we assume that for some $\eta > 0$ it holds that
\[
  \Diss(u^N;S) = \sum_{\text{$t_0$ singular point}} \Diss_{\mathrm{jump}}(u^N(t_0,\frarg);[-\eta,\eta]),
\]
where
\begin{align*}
  \Diss_{\mathrm{jump}}(u^N(t_0,\frarg);[-\eta,\eta])
  &= \int_{t_0-\eta}^{t_0+\eta} \int_\Omega \Sigma(u^N(\tau)) \circ \frac{\DD}{\DD \tau} H^N(\tau) \dd x \dd \tau  \\
  &= \int_{-\eta}^{\eta} \int_\Omega \Sigma(u^N(t_0,\theta)) \circ \frac{\DD}{\DD \theta} H^N(t_0,\theta) \dd x \dd \theta
\end{align*}
This is not rigorously true, but in the asymptotic limits $N \to \infty$, $\eta \todown 0$ it can be assumed to hold. A more detailed investigation must be postponed to the full mathematical analysis in future work since it depends on the actual, precise rescaling around singular points. Assuming the above to be true, however, we get using~\eqref{eq:Rconj_structure}, omitting the arguments $(t_0,\theta)$ for better readability,
\begin{align*}
  \Sigma(u^N) \circ \frac{\DD}{\DD \theta} H^N
    &= \frac{1}{\lambda^N} \biggl[ R \biggl( \lambda^N \frac{\DD}{\DD \tau} H^N \biggr) + R^* \bigl( \Sigma(u^N) \bigr) \biggr] \\
  &= R_1 \biggl( \frac{\DD}{\DD \tau} H^N \biggr)
    + \frac{1}{\lambda^N} R_+ \biggl( \lambda^N \frac{\DD}{\DD \tau} H^N \biggr)
    + \inf_{\Gamma \in \Scal} R_+^* \bigl( \Sigma(u^N) - \Gamma \bigr).
\end{align*}
Now, using the Fenchel inequality $R_+(\Lambda) + R_+^*(\Gamma) \geq \Lambda \circ \Gamma$, we estimate
\[
  \Sigma(u^N) \circ \frac{\DD}{\DD \theta} H^N
  \geq R_1 \biggl( \frac{\DD}{\DD \tau} H^N \biggr)
    + \inf_{\Gamma \in \Scal} \biggl( \bigl[ \Sigma(u^N) - \Gamma \bigr] \circ \frac{\DD}{\DD \tau} H^N \biggr).
\]
Thus,
\begin{align*}
  &\Diss_{\mathrm{jump}}(u^N(t_0,\frarg);[-\eta,\eta]) \\
  &\qquad\geq \Diss_1(u^N(t_0,\frarg);[-\eta,\eta]) + \Diss_{\mathrm{res}}(u^N(t_0,\frarg);[-\eta,\eta])
\end{align*}
with the \term{residual dissipation}
\[
  \Diss_{\mathrm{res}}(u^N(t_0,\frarg);[-\eta,\eta]) = \int_{-\eta}^{\eta} \int_\Omega \inf_{\Gamma \in \Scal} \biggl( \bigl[ \Sigma(u^N(t_0,\theta)) - \Gamma \bigr] \circ \frac{\DD}{\DD \tau} H^N(t_0,\theta) \biggr) \dd x \dd \theta
\]
Altogether, letting $\eta \todown 0$ (making the neighborhood around the singular points tighter and tighter), we have approximately
\begin{align*}
  \Diss(u^N;[0,T))
  &\geq \Diss_1(u^N;[0,T)) \\
  &\qquad+ \sum_{\text{$t_0$ singular point}} \bigl[ \Diss_1(u^N(t_0,\frarg)) + \Diss_{\mathrm{res}}(u^N(t_0,\frarg)) \bigr]
\end{align*}

We assume now that we may pass to the lower limit as $N\to\infty$, this can be justified by the fact that the dissipation functionals are \emph{convex} (since $R$ is convex). Then we get
\begin{align*}
  &\liminf_{N\to\infty}\, \Diss(u^N;[0,T)) \\
  &\qquad \geq \Diss_1(u_0;[0,T)) + \sum_{\text{$t_0$ singular point}} \bigl[ \Diss_1(u(t_0,\frarg)) + \Diss_{\mathrm{res}}(u(t_0,\frarg)) \bigr] \\
  &\qquad =: \Diss_+(u;[0,T)),
\end{align*}
where $\Diss_+(u;[0,T))$ is the \term{combined dissipation} of $u$ on the interval $[0,T)$. We can now pass to the lower limit in~\eqref{eq:step_energy_upper} to get the \term{upper energy inequality}
\begin{equation} \label{eq:energy_upper}
  \Ecal[T,u^N(T)] \leq \Ecal[0,u_\start] - \Diss_+(u;[0,T)) - \int_0^T \int_\Omega \dot{f}(\tau) \cdot y_0(\tau) \dd x \dd \tau.
\end{equation}

In the following we will establish the lower inequality from the stability and the minimization property. We have
\begin{align}
  &\Ecal[T,u_0(t)] - \Ecal[0,u_\start]  \notag\\
  &\qquad = \int_0^T \int_\Omega \frac{\di}{\di \tau} \Ecal[\tau,u_0(\tau)] \dd x \dd \tau  \notag\\
  &\qquad \qquad + \sum_{\text{$t_0$ singular point}} \int_{-\infty}^{+\infty} \int_\Omega \frac{\di}{\di \theta} \Ecal[t_0,u(t_0,\theta)] \dd x \dd \theta  \notag\\
  &\qquad = -\int_0^T \int_\Omega \Sigma(u_0(\tau)) \circ \frac{\DD}{\DD \tau} H_0(\tau) \dd x \dd \tau - \int_0^T \int_\Omega \dot{f}(\tau) \cdot y_0(\tau) \dd x \dd \tau  \notag\\
  &\qquad \qquad + \int_0^T \int_\Omega \bigl( - \diverg [\DD_F W(u_0(\tau))] - f(\tau) \bigr) \cdot \dot{y}_0(\tau) \dd x \dd \tau  \notag\\
  &\qquad \qquad - \sum_{\text{$t_0$ singular point}} \int_{-\infty}^{+\infty} \int_\Omega \Sigma(u(t_0,\theta)) \circ \frac{\DD}{\DD \theta} H(t_0,\theta) \dd x \dd \theta  \label{eq:lower_energy_prelim}
\end{align}
because the energy can jump over the singular points (where we have the jump transients $u(t_0,\theta)$ with $u(t_0,\pm \infty) = u_0(t_0\pm)$ as the transient). By the Euler--Lagrange equation
\begin{equation} \label{eq:cont_EL}
  - \diverg [\DD_F W(u_0(\tau,x))] - f(\tau,x) = 0,  \qquad \tau \in [0,T), \; x \in \Omega,
\end{equation}
whereby the second to last line is in fact zero.

For every regular $\tau \in [0,T)$ and $x \in \Omega$, we furthermore have $\Sigma(u_0(\tau,x)) \in \Scal = \partial R_1(0)$, which by definition means
\[
  \Sigma(u_0(\tau,x)) \circ \Lambda \leq R_1(\Lambda)  \qquad
  \text{for all $\Lambda \in \hfrak$.}
\]
In particular,
\begin{align*}
  -\int_0^t \int_\Omega \Sigma(u_0(\tau)) \circ \frac{\DD}{\DD \tau} H_0(\tau) \dd x \dd \tau
  &\geq -\int_0^t \int_\Omega R_1 \biggl( \frac{\DD}{\DD \tau} H_0(\tau) \biggr) \dd x \dd \tau \\
  &= \Diss_1(u_0,[0,t)). 
\end{align*}

On the other hand, at singular points $t_0$, for $\theta \in (-\infty,\infty)$,
\begin{align*}
  &\Sigma(u(t_0,\theta)) \circ \frac{\DD}{\DD \theta} H(t_0,\theta) \\
  &\qquad= \inf_{\Gamma \in \Scal} \biggl( \Gamma \circ \frac{\DD}{\DD \theta} H(t_0,\theta)
    + \bigl[ \Sigma(u(t_0,\theta)) - \Gamma \bigr] \circ \frac{\DD}{\DD \theta} H(t_0,\theta) \biggr) \\
  &\qquad \leq R_1 \biggl( \frac{\DD}{\DD \theta} H(t_0,\theta) \biggr)
    + \inf_{\Gamma \in \Scal} \biggl( \bigl[ \Sigma(u^N(t_0,\theta)) - \Gamma \bigr] \circ \frac{\DD}{\DD \theta} H(t_0,\theta) \biggr).
\end{align*}
Thus,
\begin{align}
  &-\int_{-\infty}^{+\infty} \int_\Omega \Sigma(u(t_0,\theta)) \circ \frac{\DD}{\DD \theta} H(t_0,\theta) \dd x \dd \theta  \notag\\
  &\qquad\geq -\Diss_1(u(t_0,\frarg)) + \Diss_{\mathrm{res}}(u(t_0,\frarg))  \label{eq:cont_sing}
\end{align}

Using~\eqref{eq:cont_EL}--\eqref{eq:cont_sing}  in~\eqref{eq:lower_energy_prelim}, we arrive at the \term{lower energy inequality}
\[
  \Ecal[T,u^N(T)] \geq \Ecal[0,u_\start] - \Diss_+(u;[0,T)) - \int_0^T\int_\Omega \dot{f}(\tau) \cdot y_0(\tau) \dd x \dd \tau.
\]
Therefore, combining with~\eqref{eq:energy_upper} and realizing that the same argument also works with any regular $t \in [0,T)$ instead of $T$, we have finally shown the \term{energy balance}
\begin{enumerate}
  \item[(E3)]\quad $\displaystyle
  \Ecal[t,u^N(t)] = \Ecal[0,u_\start] - \Diss_+(u;[0,t)) - \int_0^t \int_\Omega \dot{f}(\tau) \cdot y_0(\tau) \dd x \dd \tau.$
\end{enumerate}

\subsection{Rate-independent flow rule}

Differentiating the the energy balance~(E3) with respect to time at a regular point $t \in [0,T)$ (away from any singular points) and dropping the integral (the following argument works on any subdomain $\Omega' \subset \Omega$), we get
\[
  -\Sigma(u_0(t)) \circ \frac{\DD}{\DD t} H_0(t)
    + \bigl( -\diverg [\DD_F W(u_0(t))] - f(t) \bigr) \cdot \dot{y}_0(t)
  = -R_1 \biggl( \frac{\DD}{\DD t} H_0(t) \biggr).
\]
Thus, also employing the Euler--Lagrange equation~\eqref{eq:cont_EL}, we get
\[
  R_1 \biggl( \frac{\DD}{\DD t} H_0(t) \biggr) - \Sigma(u_0(t)) \circ \frac{\DD}{\DD t} H_0(t)
  = 0.
\]
Additionally, we have from the stability~(E2) that $\Sigma(u_0(t,x)) \in \Scal = \partial R_1(0)$, i.e.\
\[
  \Sigma(u_0(t)) \circ \Lambda \leq R_1(\Lambda)  \qquad
  \text{for all $\Lambda \in \hfrak$.}
\]
Adding the last two assertions, we have
\[
  R_1 \biggl( \frac{\DD}{\DD t} H_0(t) \biggr) + \Sigma(u_0(t)) \circ \biggl( \Lambda - \frac{\DD}{\DD t} H_0(t) \biggr) \leq R_1(\Lambda)  \qquad
  \text{for all $\Lambda \in \hfrak$.}
\]
This is nothing else than the written-out formula of the \term{rate-independent differential inclusion}
\begin{enumerate}
  \item[(E4)]\quad $\displaystyle
  \Sigma(u_0(t,x)) \in \partial R_1 \biggl( \frac{\DD}{\DD t} H_0(t,x) \biggr)
  \qquad\text{for all regular $t \in [0,T)$, $x \in \Omega$.}$
\end{enumerate}

%

\subsection{Two-speed solutions}

\begin{figure}[tb]
\begin{center}
\includegraphics[scale=0.9]{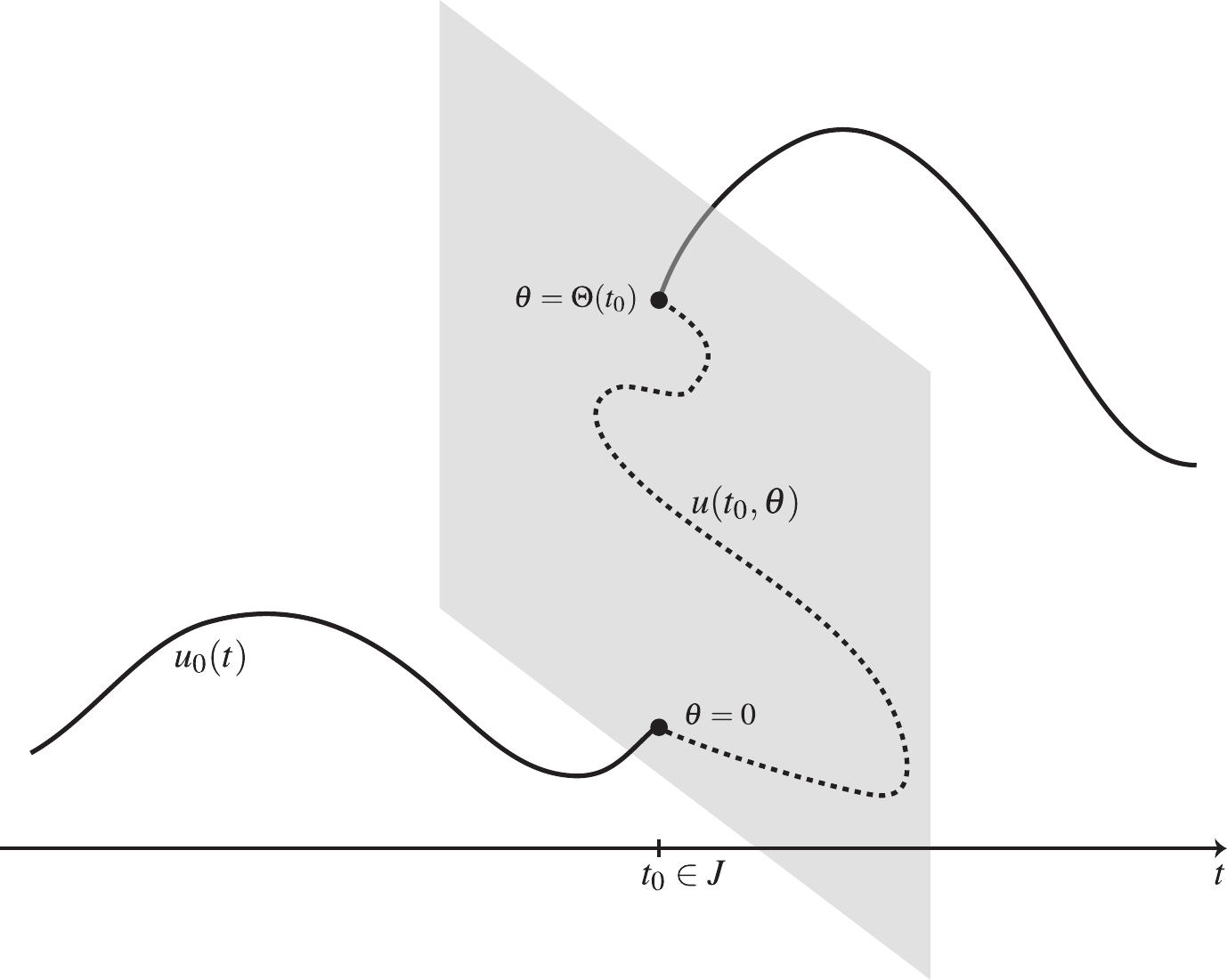}
\caption{A two-speed solution.} 
\label{fig:twospeed}
\end{center}
\end{figure}

A process $u_0 = (y_0,H_0) \colon [0,T) \times \Omega \to \R^d \times \Hfrak$ together with \emph{transients}
\[
  u(t_0,\frarg,\frarg) = (y_0(t_0,\frarg,\frarg), H_0(t_0,\frarg,\frarg)) \colon (-\infty,\infty) \times \Omega \to \R^d \times \Hfrak
\]
with
\[
  u(t_0,\pm \infty) = u_0(t_0\pm)
\]
for all \term{singular (jump) points} $t_0 \in J \subset [0,T)$ is called a \term{two-speed solution} if the following assertions are true:
\begin{enumerate}
  \item[(E1$_{\mathrm{reg}}$)]\quad $\displaystyle
  y_0(t_0) \quad \text{minimizes} \quad
  \hat{y} \mapsto \Ecal[t_0,\hat{y},H_0(t_0)]  \qquad
  \text{for regular $t_0 \in [0,T)$.}$
  
  \item[(E1$_{\mathrm{sing}}$)]\quad $\displaystyle
  y(t_0,\theta) \quad \text{minimizes} \quad
  \hat{y} \mapsto \Ecal[t_0,\hat{y},H(t_0,\theta)]  \qquad
  \text{for singular $t_0 \in [0,T)$.}$
  
  \item[(E2)]\quad $\displaystyle
  \Sigma(u_0(t_0,x)) \in \Scal \quad\text{for $t_0$ regular, $x \in \Omega$.}$
  
  \item[(E3)]\quad $\displaystyle
  \Ecal[t,u^N(t)] = \Ecal[0,u_\start] - \Diss_+(u;[0,t)) - \int_0^t \int_\Omega \dot{f}(\tau,x) \cdot y_0(\tau,x) \dd x \dd \tau$ for all regular $t \in [0,T)$.
 
  \item[(E4)]\quad $\displaystyle
  \Sigma(u_0(t,x)) \in \partial R_1 \biggl( \frac{\DD}{\DD t} H_0(t,x) \biggr)
  \qquad\text{for all regular $t \in [0,T)$, $x \in \Omega$.}$
\end{enumerate}
An illustration of a two-speed solution is in Figure~\ref{fig:twospeed} (here, $\theta$ in $u(t_0,\theta)$ runs only in the interval $[0,\Theta(t_0)]$, i.e.\ the jump length is finite).

\section{Conclusion}

The present manuscript aims to make a contribution to the mathematical modelling of large-strain elasto-plasticity and to provide a conceptual framework for existence theorems. Future work will cast the concepts introduced into this work, in particular the two-speed solutions, into a mathematically rigorous framework and perform a thorough mathematical analysis.

\subsection*{Acknowledgements}

I would like to thank Sergio Conti, Georg Dolzmann, Gilles Francfort, Michael Ortiz, Ulisse Stefanelli, Florian Theil, and Emil Wiedemann (in particular, Example~\ref{ex:D_nonconvex} is due to him) for many interesting discussions related to the present work. The support of the author through the EPSRC Research Fellowship EP/L018934/1 on \enquote{Singularities in Nonlinear PDEs} and through the Royal Society Travel Grant IE131532 is gratefully acknowledged.

\subsection*{Author's biography}

Filip Rindler has held the post of Zeeman Lecturer (Assistant Professor) in Mathematics at the University of Warwick since 2013. After completing his doctorate in nonlinear PDE theory and the calculus of variations at the OxPDE centre within the University of Oxford in 2011, he moved to the University of Cambridge to take up the Gonville \& Caius College Drosier Research Fellowship, holding the position until 2015 (on leave 2013--2015). For 2014--2017 he is funded by a full-time EPSRC Research Fellowship on \enquote{Singularities in Nonlinear PDEs}.


\begin{thebibliography}{10}

\bibitem{AbbaschianReedHill09}
R.~Abbaschian, L.~Abbaschian, and R.~E. Reed-Hill, \emph{{Physical Metallurgy
  Principles - SI Edition}}, Cengage Learning, 2009.

\bibitem{AmbrosioFuscoPallara00}
L.~Ambrosio, N.~Fusco, and D.~Pallara, \emph{{Functions of Bounded Variation
  and Free-Discontinuity Problems}}, Oxford Mathematical Monographs, Oxford
  University Press, 2000.

\bibitem{ArmstrongArnoldZerilli09}
R.~W. Armstrong, W.~Arnold, and F.~J. Zerilli, \emph{Dislocation mechanics of
  copper and iron in high rate deformation tests}, J. Appl. Phys. \textbf{105}
  (2009), 023511.

\bibitem{BenDavidEtAl14}
E.~Ben-David, T.~Tepper-Faran, D.~Rittel, and D.~Shilo, \emph{A large strain
  rate effect in thin free-standing {Al} films}, Scripta Materialia
  \textbf{90--91} (2014), 6--9.

\bibitem{CaseyNaghdi80}
J.~Casey and P.~M. Naghdi, \emph{{A Remark on the Use of the Decomposition F =
  FeFp in Plasticity}}, J. Appl. Mech. \textbf{47} (1980), 672--675.

\bibitem{Ciarlet88}
P.~Ciarlet, \emph{{Mathematical Elasticity}}, vol. 1: Three Dimensional
  Elasticity, North-Holland, 1988.

\bibitem{ColemanNoll63}
B.~D. Coleman and W.~Noll, \emph{{The thermodynamics of elastic materials with
  heat conduction and viscosity}}, Arch. Ration. Mech. Anal. \textbf{13}
  (1963), no.~1, 167--178.

\bibitem{ContiTheil05}
S.~Conti and F.~Theil, \emph{{Single-Slip Elastoplastic Microstructures}},
  Arch. Ration. Mech. Anal. \textbf{178} (2005), 125--148.

\bibitem{Dafalias87}
Y.~F. Dafalias, \emph{{Issues on the constitutive formulation at large
  elastoplastic deformations, part 1: Kinematics}}, Acta Mech. \textbf{69}
  (1987), 119--138.

\bibitem{DalMasoDeSimoneSolombrino10}
G.~Dal~Maso, A.~DeSimone, and F.~Solombrino, \emph{{Quasistatic evolution for
  Cam-Clay plasticity: a weak formulation via viscoplastic regularization and
  time rescaling}}, Calc. Var. Partial Differential Equations \textbf{40}
  (2010), 125--181.

\bibitem{DalMasoDeSimoneSolombrino11}
\bysame, \emph{{Quasistatic evolution for Cam-Clay plasticity: properties of
  the viscosity solution}}, Calculus of Variations and Partial Differential
  Equations \textbf{44} (2011), 495--541.

\bibitem{DeHossonyRoosMetselaar02}
T.~M. {De Hossony}, A.~Roos, and E.~D. Metselaar, \emph{Temperature rise due to
  fast-moving dislocations}, Philos. Mag. A \textbf{81} (2001), 1099--1120.

\bibitem{EvansGariepy92}
L.~C. Evans and R.~F. Gariepy, \emph{{Measure Theory and Fine Properties of
  Functions}}, Studies in Advanced Mathematics, CRC Press, 1992.

\bibitem{FrancfortMielke06}
G.~A. Francfort and A.~Mielke, \emph{Existence results for a class of
  rate-independent material models with nonconvex elastic energies}, J. Reine
  Angew. Math. \textbf{595} (2006), 55--91.

\bibitem{Fremond02}
M.~Fr\'{e}mond, \emph{{Non-Smooth Thermomechanics}}, Springer, 2002.

\bibitem{GrandiStefanelli15}
D.~Grandi and U.~Stefanelli, \emph{{Finite plasticity in P$^{T}$ P}},
  arXiv:1509.08681, 2015.

\bibitem{GreenNaghdi71}
A.~E. Green and P.~M. Naghdi, \emph{{Some remarks on elastic-plastic
  deformation at finite strain}}, International Journal of Engineering Science
  \textbf{9} (1971), 1219--1229.

\bibitem{Gurtin2000}
M.~E. Gurtin, \emph{{On the plasticity of single crystals: free energy,
  microforces, plastic-strain gradients}}, J. Mech. Phys. Solids \textbf{48}
  (2000), 989--1036.

\bibitem{GurtinFriedAnand10}
M.~E. Gurtin, E.~Fried, and L.~Anand, \emph{The mechanics and thermodynamics of
  continua}, Cambridge University Press, 2010. \MR{2884384 (2012m:74001)}

\bibitem{Hall03}
B.~C. Hall, \emph{Lie groups, {L}ie algebras, and representations}, Graduate
  Texts in Mathematics, vol. 222, Springer, 2003.

\bibitem{HanReddy13}
W.~Han and B.~D. Reddy, \emph{{Plasticity -- Mathematical Theory and Numerical
  Analysis}}, Interdisciplinary Applied Mathematics, vol.~9, Springer, 2013.

\bibitem{HochrainerEtAl14}
T.~Hochrainer, S.~Sandfeld, M.~Zaiser, and P.~Gumbsch, \emph{{Continuum
  dislocation dynamics: Towards a physical theory of crystal plasticity}},
  Journal of the Mechanics and Physics of Solids \textbf{63} (2014), 167--178.

\bibitem{HochrainerZaiserGumbsch07}
T~Hochrainer, M~Zaiser, and P~Gumbsch, \emph{{A three-dimensional continuum
  theory of dislocation systems: kinematics and mean-field formulation}},
  Philos. Mag. \textbf{87} (2007), 1261--1282.

\bibitem{KassnerEtAl09}
M.~E. Kassner, P.~Geantil, L.~E. Levine, and B.~C. Larson, \emph{Backstress,
  the {Bauschinger} effect and cyclic deformation}, Mater. Sci. Forum
  \textbf{604--605} (2009), 39--51.

\bibitem{Kroner60}
E.~Kr{\"o}ner, \emph{Allgemeine {K}ontinuumstheorie der {V}ersetzungen und
  {E}igenspannungen}, Arch. Ration. Mech. Anal. (1960), 273--334.

\bibitem{Kroner01}
\bysame, \emph{{Benefits and shortcomings of the continuous theory of
  dislocations}}, Int. J. Solids Struct. \textbf{38} (2001), 1115--1134.

\bibitem{Lee69EPDF}
E.~H. Lee, \emph{Elastic-plastic deformation at finite strain}, ASME J. Appl.
  Mech. \textbf{36} (1969), 1--6.

\bibitem{LeeLiu67FSEP}
E.~H. Lee and D.~T. Liu, \emph{Finite-strain elastic-plastic theory with
  application to plane-wave analysis}, J. Appl. Phys. \textbf{38} (1967),
  19--27.

\bibitem{Lee13}
J.~M. Lee, \emph{{Introduction to Smooth Manifolds}}, 2nd ed., Graduate Texts
  in Mathematics, vol. 218, Springer, 2013.

\bibitem{LemaitreChaboche90}
J.~Lemaitre and J.-L. Chaboche, \emph{Mechanics of solid materials}, Cambridge
  University Press, 1990.

\bibitem{LubardaLee81}
V.~A. Lubarda and E.~H. Lee, \emph{{A Correct Definition of Elastic and Plastic
  Deformation and Its Computational Significance}}, J. Appl. Mech. \textbf{48}
  (1981), 35, Erratum: J. Appl. Mech. 48 (1981), 686.

\bibitem{Lubliner08}
J.~Lubliner, \emph{{Plasticity Theory}}, Dover, 2008.

\bibitem{MainikMielke05}
A.~Mainik and A.~Mielke, \emph{Existence results for energetic models for
  rate-independent systems}, Calc. Var. Partial Differential Equations
  \textbf{22} (2005), 73--99.

\bibitem{MainikMielke09}
A.~Mainik and A.~Mielke, \emph{Global existence for rate-independent gradient
  plasticity at finite strain}, J. Nonlinear Sci. \textbf{19} (2009), 221--248.

\bibitem{Mandel73}
J.~Mandel, \emph{{Equations constitutives et directeurs dans les milieux
  plastiques et viscoplastiques}}, Int. J. Solids Struct. \textbf{9} (1973),
  725--740.

\bibitem{Mielke02}
A.~Mielke, \emph{Finite elastoplasticity {L}ie groups and geodesics on {${\rm
  SL}(d)$}}, Geometry, mechanics, and dynamics, Springer, 2002, pp.~61--90.

\bibitem{Mielke03a}
\bysame, \emph{{Energetic formulation of multiplicative elasto-plasticity using
  dissipation distances}}, Contin. Mech. Thermodyn. \textbf{15} (2003),
  351--382.

\bibitem{MielkeMuller06}
A.~Mielke and S.~M{\"u}ller, \emph{Lower semicontinuity and existence of
  minimizers in incremental finite-strain elastoplasticity}, ZAMM Z. Angew.
  Math. Mech. \textbf{86} (2006), 233--250.

\bibitem{MielkeRindler09}
A.~Mielke and F.~Rindler, \emph{Reverse approximation of energetic solutions to
  rate-independent processes}, NoDEA Nonlinear Differential Equations Appl.
  \textbf{16} (2009), 17--40.

\bibitem{MielkeRossiSavare09}
A.~Mielke, R.~Rossi, and G.~Savar\'{e}, \emph{Modeling solutions with jumps for
  rate-independent systems on metric spaces}, Discrete Contin. Dyn. Syst.
  \textbf{25} (2009), 585--615.

\bibitem{MielkeRossiSavare12}
\bysame, \emph{B{V} solutions and viscosity approximations of rate-independent
  systems}, ESAIM Control Optim. Calc. Var. \textbf{18} (2012), 36--80.

\bibitem{MielkeRoubicek15book}
A.~Mielke and T.~Roub\'{i}\v{c}ek, \emph{{Rate-Independent Systems: Theory and
  Application.}}, Applied Mathematical Sciences, vol. 193, Springer, 2015.

\bibitem{MielkeTheil99}
A.~Mielke and F.~Theil, \emph{A mathematical model for rate-independent phase
  transformations with hysteresis}, Proceedings of the Workshop on ``Models of
  Continuum Mechanics in Analysis and Engineering'' (H.-D. Alber, R.M. Balean,
  and R.~Farwig, eds.), Shaker Verlag, 1999, pp.~117--129.

\bibitem{MielkeTheil04}
\bysame, \emph{On rate-independent hysteresis models}, NoDEA Nonlinear
  Differential Equations Appl. \textbf{11} (2004), 151--189.

\bibitem{MielkeTheilLevitas02}
A.~Mielke, F.~Theil, and V.~I. Levitas, \emph{A variational formulation of
  rate-independent phase transformations using an extremum principle}, Arch.
  Ration. Mech. Anal. \textbf{162} (2002), 137--177.

\bibitem{Naghdi90}
P.~M. Naghdi, \emph{{A critical review of the state of finite plasticity}},
  ZAMP Zeitschrift f\"{u}r angewandte Mathematik und Physik \textbf{41} (1990),
  315--394.

\bibitem{NematNasser79}
S.~Nemat-Nasser, \emph{{Decomposition of strain measures and their rates in
  finite deformation elastoplasticity}}, Int. J. Solids Struct. \textbf{15}
  (1979), 155--166.

\bibitem{OrtizRepetto99}
M.~Ortiz and E.~A. Repetto, \emph{Nonconvex energy minimization and dislocation
  structures in ductile single crystals}, J. Mech. Phys. Solids \textbf{47}
  (1999), 397--462.

\bibitem{ReinaConti14}
C.~Reina and S.~Conti, \emph{{Kinematic description of crystal plasticity in
  the finite kinematic framework: A micromechanical understanding of F=FeFp}},
  J. Mech. Phys. Solids \textbf{67} (2014), 40--61.

\bibitem{Rice71}
J.~R. Rice, \emph{{Inelastic constitutive relations for solids: An
  internal-variable theory and its application to metal plasticity}}, J. Mech.
  Phys. Solids \textbf{19} (1971), 433--455.

\bibitem{Rindler15TwoSpeedRI}
F.~Rindler, \emph{Two-speed solutions for rate-independent systems},
  forthcoming, 2015.

\bibitem{Rock70CA}
R.~T. Rockafellar, \emph{{Convex Analysis}}, Princeton Mathematical Series,
  vol.~28, Princeton University Press, 1970.

\bibitem{SandfeldEtAl11}
S.~Sandfeld, T.~Hochrainer, M.~Zaiser, and P.~Gumbsch, \emph{{Continuum
  modeling of dislocation plasticity: Theory, numerical implementation, and
  validation by discrete dislocation simulations}}, J. Mater. Res. \textbf{26}
  (2011), 623--632.

\bibitem{Silhavy97}
M.~Silhavy, \emph{{The Mechanics and Thermodynamics of Continuous Media}},
  Springer, 1997.

\bibitem{Stefanelli08}
U.~Stefanelli, \emph{{A Variational Principle for Hardening Elastoplasticity}},
  SIAM J. Math. Anal. \textbf{40} (2008), 623--652.

\bibitem{Stefanelli09}
\bysame, \emph{{A variational characterization of rate-independent evolution}},
  Math. Nachr. \textbf{282} (2009), 1492--1512.

\bibitem{Zbib93}
H.~M. Zbib, \emph{{On the mechanics of large inelastic deformations: kinematics
  and constitutive modeling}}, Acta Mechanica \textbf{96} (1993), 119--138.

\bibitem{ZieglerWehrli87}
H.~Ziegler and C.~Wehrli, \emph{The derivation of constitutive relations from
  the free energy and the dissipation function}, Advances in applied mechanics,
  Vol.\ 25, Adv. Appl. Mech., vol.~25, Academic Press, 1987, pp.~183--237.

\end{thebibliography}


\providecommand{\bysame}{\leavevmode\hbox to3em{\hrulefill}\thinspace}
\providecommand{\MR}{\relax\ifhmode\unskip\space\fi MR }
\providecommand{\MRhref}[2]{%
  \href{http://www.ams.org/mathscinet-getitem?mr=#1}{#2}
}
\providecommand{\href}[2]{#2}

\end{document}